\documentclass[a4paper,12pt]{article}

\usepackage{amsmath}
\usepackage{amssymb}
\usepackage[dvips]{graphicx}
\usepackage[dvips]{psfrag}
\usepackage{bm}

\newcommand{\id}{{\bf 1}} 
\newcommand {\be}{\begin{equation}}
\newcommand {\ee}{\end{equation}}
\newcommand {\bea}{\begin{eqnarray}}
\newcommand {\eea}{\end{eqnarray}}
\newcommand {\nn}{\nonumber}
\newcommand {\tr}{{\rm tr}}

\newcommand{\cO}{{\cal O}}

\newcommand{\vev}[1]{\left\langle #1 \right\rangle}

\begin{document}
\thispagestyle{empty} \addtocounter{page}{-1}
\begin{flushright}
OIQP-13-11,\, RIKEN-QHP-89
%
\end{flushright} 
\vspace*{5mm}

\begin{center}
{\large \bf SUSY breaking by nonperturbative dynamics}\\
{\large \bf in a matrix model for 2D type IIA superstrings}\\
\vspace*{1cm}
Michael G. Endres$^*$, Tsunehide Kuroki$^\dagger$, Fumihiko Sugino$^\ddagger$ and Hiroshi Suzuki$^*$\\
\vskip0.7cm
{}$^*${\it Theoretical Research Division, RIKEN Nishina Center,}\\
\vspace*{1mm} 
{\it Wako, Saitama 351-0198, Japan}\\
\vspace*{0.2cm}
{\tt endres@riken.jp}, \, {\tt hsuzuki@riken.jp}\\
\vskip0.4cm 
{}$^\dagger${\it Kobayashi-Maskawa Institute for the Origin of Particles and the Universe, }\\
\vspace*{1mm}
{\it Nagoya University, Nagoya 464-8602, Japan}\\
\vspace*{0.2cm}
{\tt kuroki@kmi.nagoya-u.ac.jp}\\
\vskip0.4cm
{}$^\ddagger${\it Okayama Institute for Quantum Physics, } \\
\vspace*{1mm}
{\it Kyoyama 1-9-1, Kita-ku, Okayama 700-0015, Japan}\\
\vspace*{0.2cm}
{\tt fumihiko\_sugino@pref.okayama.lg.jp}\\
\end{center}
\vskip2cm
\centerline{\bf Abstract}
\vspace*{0.3cm}
{\small 
We explicitly compute nonperturbative effects in a supersymmetric double-well matrix model
corresponding to two-dimensional type IIA superstring theory on a nontrivial Ramond-Ramond 
background. We analytically determine the full one-instanton contribution to the free energy and
one-point function, including all perturbative fluctuations around 
the one-instanton background. The leading order two-instanton contribution is determined as well. 
We see that 
supersymmetry is spontaneously broken by instantons, and that 
the breaking persists after taking a double scaling limit which realizes the type IIA theory from the 
matrix model. The result implies that spontaneous supersymmetry breaking occurs by nonperturbative dynamics 
in the target space of the IIA theory.  
Furthermore, we numerically determine the full nonperturbative effects by recursive evaluation of orthogonal polynomials.
The free energy of the matrix model appears well-defined and finite even in the strongly 
coupled limit of the corresponding type IIA theory. 
The result might suggest a weakly coupled theory appearing 
as an S-dual to the two-dimensional type 
IIA superstring theory.  
}
\vspace*{1.1cm}



\newpage

\section{Introduction}
Matrix models for noncritical string theory have been vigorously investigated as toy models 
for critical string theory since the late 1980s, 
and have unveiled interesting nonperturbative structures behind the theory 
(for reviews, see~\cite{Di Francesco:1993nw,Klebanov:1991qa,Ginsparg:1993is}). 
More recently, these matrix models have been understood from the perspective of decaying 
D-branes~\cite{McGreevy:2003kb,Klebanov:2003km,McGreevy:2003ep,Alexandrov:2003nn,Takayanagi:2003sm,Douglas:2003up,McGreevy:2003dn}, 
although such discussions have been primarily confined to 
bosonic string theory or superstring theory without target-space supersymmetry. 
Since little is known about 
(solvable) matrix models corresponding to noncritical superstrings with target-space supersymmetry, 
it would be an important direction to find such matrix models and investigate their nonperturbative properties. 
Particularly these are expected to possess aspects more relevant for critical superstring theory compared with 
the matrix models for string theory without target-space supersymmetry. 

As a step along this direction, 
a supersymmetric double-well matrix model had recently been considered in zero dimensions, 
and its connection to two-dimensional type IIA superstring theory 
on a nontrivial Ramond-Ramond background had been explored from the viewpoint of symmetries and the spectrum~\cite{Kuroki:2012nt}. 
The target space of the type IIA theory is 
$(\varphi, x)\in \mbox{(Liouville direction)} \times \mbox{($S^1$ with self-dual radius)}$, 
and the holomorphic energy-momentum tensor on the string worldsheet (excluding the ghost part) is given by 
\be
T_{\rm m} = -\frac12 (\partial x)^2 -\frac12 \psi_x\partial\psi_x -\frac12(\partial \varphi)^2 
+\partial^2\varphi -\frac12\psi_\ell\partial \psi_\ell, 
\label{EMtensor_IIA}
\ee
where $\psi_x$ and $\psi_\ell$ are superpartners of $x$ and $\varphi$, respectively. 
The anti-holomorphic energy-momentum tensor has a similar expression. 
Target-space supercharges are represented by contour integrals of vertex operators: 
\be
Q_+=\oint \frac{dz}{2\pi i}\,e^{-\frac12\phi-\frac{i}{2}H-ix}(z), \qquad 
\bar{Q}_-= \oint \frac{d\bar{z}}{2\pi i}\,e^{-\frac12\bar{\phi}+\frac{i}{2}\bar{H}+i\bar{x}}(\bar{z})
\label{supercharge_IIA}
\ee
with cocycle factors suppressed. $\phi$ denotes the bosonized superconformal ghost, and $H$ is a scalar field 
introduced for bosonization: $\psi_\ell\pm i\psi_x=\sqrt{2}\,e^{\mp i H}$. 
The field variables with bars belong to the anti-holomorphic sector. The supercharges satisfy 
\be
Q_+^2=\bar{Q}_-^2=\{ Q_+, \,\bar{Q}_-\}=0. 
\label{nilpotency_IIA}
\ee
The Ramond-Ramond background preserves the supersymmetry, because the corresponding 
Ramond-Ramond vertex operators transform as singlets under the supersymmetry. 
In addition, dynamical aspects of the connection were explicitly shown in~\cite{Kuroki:2013qpa} 
by comparing scattering amplitudes computed in the matrix model with those in the type IIA theory. 
The comparison was mainly made between correlation functions 
in a normal large-$N$ limit (planar limit) of the matrix model and tree amplitudes in the type IIA 
superstring theory. 
 
In this paper, we calculate nonperturbative effects of the supersymmetric double-well matrix model 
and discuss their implications in the corresponding two-dimensional type IIA superstring theory. 
In order to realize the type IIA theory from the matrix model beyond tree level, 
one should take a double scaling limit 
which sends the size of the matrices $N$ to infinity with the coupling constant $\mu$ approaching 
a critical value at an appropriate $N$-dependent rate 
(for example, see~\cite{Brezin:1990rb,Douglas:1989ve,Gross:1989aw}). 
In the type IIA theory, 
the limit corresponds to taking into account each order of the string perturbation series on an equal footing 
and incorporating nonperturbative effects.  
The full one-instanton contribution to the free energy and one-point function,
including all perturbative fluctuations around 
the one-instanton background, is obtained in the double scaling limit. 
The leading two-instanton contribution is determined in this limit as well~\footnote{
In this paper, the leading $k$-instanton contribution means the leading order term of the $k$-instanton 
contribution. This term contains a contribution 
from the classical 
$k$-instanton configuration as well as from the one-loop fluctuations around the $k$-instanton 
background.}. 
We find that the instanton effects break the supersymmetry of the model, and  
that the breaking survives in the double scaling limit. 
The result is remarkable since in a simple large-$N$ limit (with $\mu$ fixed) 
supersymmetry breaking by instantons ceases 
and the supersymmetry becomes restored~\cite{Witten:1982df,Affleck:1983pn}~\footnote{
Some ways around the issue are discussed in~\cite{Kuroki:2007iy,Kuroki:2009yg}.}. 
Moreover, we numerically determine the full nonperturbative effects, which 
give further evidence of supersymmetry breaking in the double scaling limit. 
Thus, our supersymmetric matrix model provides a valuable framework 
for describing a superstring theory whose target-space supersymmetry is broken by nonperturbative dynamics. 
It would be intriguing to consider matrix models for critical superstring theory 
exhibiting similar properties, as these may be interesting candidates for describing the real world. 

The rest of this paper is organized as follows. 
In the next section, the supersymmetric double-well matrix model is introduced, and its partition function 
is regularized in order to define an order parameter for spontaneous supersymmetry breaking. 
In section~\ref{sec:MMinst}, we compute the leading one-instanton contribution 
to the partition function following an 
approach used for the $c=0$ matrix model discussed in~\cite{Hanada:2004im,Ishibashi:2005dh}. 
We introduce orthogonal polynomials for our matrix model in section~\ref{sec:ortho_poly}, 
and then make use of them in sections~\ref{sec:1-inst} and \ref{sec:2-inst}
to compute nonperturbative effects in a more efficient way.
In particular, the full one-instanton contribution including all perturbative fluctuations 
around the one-instanton background is obtained in section~\ref{sec:1-inst_full}, 
and the leading two-instanton contribution is obtained in section~\ref{sec:2-inst}. 
We show from these results that supersymmetry is spontaneously broken even after taking the double scaling limit. 
Note that we do not use the dilute gas approximation for instantons, 
and that interactions among instantons are taken into account. 
In section~\ref{sec:full}, we numerically calculate the orthogonal polynomials using Mathematica in order to 
evaluate the full nonperturbative effects. Interestingly, the free energy seems to be well-defined and finite 
even in the strongly coupled limit of the corresponding type IIA theory. 
This might suggest a weakly coupled theory appearing as an S-dual to the two-dimensional type IIA superstring 
theory. 
In section~\ref{sec:discussions}, we summarize the results obtained thus far, and discuss 
some future directions. 
Appendix~\ref{app:localization} is devoted to a perturbative calculation of the partition function 
by a deformation method used in topological field theory. 
In appendix~\ref{app:Veff1}, we present a computation of the effective potential for a single eigenvalue 
at subleading order in 
$1/N$, which is necessary for evaluating the leading one-instanton effect in section~\ref{sec:MMinst}. 
An asymptotic formula for the Hermite polynomials required to obtain the full one-instanton contribution 
is derived in appendix~\ref{app:formula}.
Finally, we present a plot for results at subleading order in large $N$ in the double scaling limit 
in appendix~\ref{app:1ptfn_subleadingN}.

\section{A supersymmetric double-well matrix model}
\label{sec:SUSYMM}
\setcounter{equation}{0}
The action and partition function for the supersymmetric double-well matrix model 
introduced in~\cite{Kuroki:2012nt,Kuroki:2009yg,Kuroki:2010au} are 
given by 
\be
S = N \tr \left[\frac12 B^2 +iB(\phi^2-\mu^2) +\bar\psi (\phi\psi+\psi\phi)\right] 
\label{S}
\ee
and
\be
Z =  \left(-1\right)^{N^2}
\int d^{N^2}B \,d^{N^2}\phi \,\left(d^{N^2}\psi \,d^{N^2}\bar{\psi}\right)\, e^{-S},  
\label{Z}
\ee
where $B$ and $\phi$ are $N\times N$ Hermitian matrices, and $\psi$ and $\bar\psi$ are $N\times N$ 
Grassmann-odd matrices. We fix the normalization of the measure such that
\be
\int d^{N^2}\phi \, e^{-N\tr \,(\frac12 \phi^2)} = \int d^{N^2}B \, e^{-N\tr \,(\frac12 B^2)} = 1
\label{normalization1}
\ee
and 
\be 
(-1)^{N^2} \int \left(d^{N^2}\psi \,d^{N^2}\bar{\psi}\right)\, e^{-N\tr \,(\bar{\psi}\psi)}=1. 
\label{normalization2}
\ee
The coupling constant $\mu$ is considered in this work to be real and positive. 
The action $S$ is invariant under supersymmetry transformations generated by $Q$ and $\bar{Q}$, given by: 
\be
Q\phi =\psi, \quad Q\psi=0, \quad Q\bar{\psi} =-iB, \quad QB=0, 
\label{QSUSY}
\ee
and 
\be
\bar{Q} \phi = -\bar{\psi}, \quad \bar{Q}\bar{\psi} = 0, \quad 
\bar{Q} \psi = -iB, \quad \bar{Q} B = 0,  
\label{QbarSUSY}
\ee
which leads to the nilpotency: $Q^2=\bar{Q}^2=\{ Q, \bar{Q}\}=0$. 
This is isomorphic to (\ref{nilpotency_IIA}) in the type IIA superstring theory. 
Furthermore, by comparing (\ref{QSUSY}) and (\ref{QbarSUSY}) with the $Q_+$ and $\bar{Q}_-$ transformations of 
vertex operators in the type IIA theory and computing scattering amplitudes in both sides, the correspondence
\be
(Q, \bar{Q}) \Leftrightarrow (Q_+, \bar{Q}_-)
\ee 
is confirmed between the matrix model and the type IIA theory~\cite{Kuroki:2012nt,Kuroki:2013qpa}. 

After integrating out all matrices other than $\phi$, the partition function (\ref{Z}) is expressed as 
\bea
Z & = & \int d^{N^2}\phi\,e^{-N\frac12\tr(\phi^2-\mu^2)^2}\,\det(\phi\otimes \id +\id\otimes \phi) \nn \\
 & = & \tilde{C}_N\int \Big(\prod_{i=1}^Nd\lambda_i\Big)\,\triangle(\lambda)^2\,
\prod_{i,j=1}^N(\lambda_i+\lambda_j)\,e^{-N\sum_{i=1}^N\frac12(\lambda_i^2-\mu^2)^2}, 
\label{Zeigen}
\eea
where $\id$ is an $N\times N$ unit matrix. 
In the last line, the expression reduces to integrals with respect to the $N$ eigenvalues $\lambda_i$ ($i=1,\ldots,N$) of $\phi$. $\triangle(\lambda)$ 
denotes the Vandermonde determinant $\triangle(\lambda)=\prod_{i>j}(\lambda_i-\lambda_j)$, and 
$\tilde{C}_N$ is a numerical factor depending only on $N$ given by 
\be
\frac{1}{\tilde{C}_N} = \int \Bigl(\prod_{i=1}^N d\lambda_i\Bigr) \,\triangle(\lambda)^2\, 
e^{-N\sum_{i=1}^N \frac12 \lambda_i^2}
=(2\pi)^{\frac{N}{2}}\frac{\prod_{k=0}^Nk!}{N^{\frac{N^2}{2}}}.
\label{CtildeN}
\ee   
In this paper, we work with the partition function in the sector with filling fraction~\footnote{
$\nu_{\pm}$ are nonnegative fractional numbers such that $\nu_++\nu_-=1$ and $\nu_{\pm}N$ are integers.}
 $(\nu_+, \nu_-)$ 
which is defined by  
\bea
Z_{(\nu_+,\nu_-)} & \equiv & \tilde{C}_N\int_0^{\infty}\left(\prod_{i=1}^{\nu_+N}d\lambda_i\right) 
\int_{-\infty}^0\left(\prod_{j=\nu_+N+1}^N d\lambda_j\right)\, \left(\prod_{n=1}^N2\lambda_n\right)\,
\left\{\prod_{n>m}(\lambda_n^2-\lambda_{m}^2)^2\right\}\nn \\
& & \hspace{68mm}\times e^{-N\sum_{i=1}^N\frac12(\lambda_i^2-\mu^2)^2}. 
\label{Zff}
\eea
The integration region 
of each eigenvalue is divided into the positive and negative 
real axes. $Z_{(\nu_+,\nu_-)}$ represents the part of $Z$, where the first $\nu_+N$ eigenvalues are 
integrated over the positive real axis and the remaining $\nu_-N$ are integrated over the negative real axis. 
Intuitively, the dominant contribution to 
$Z_{(\nu_+,\nu_-)}$ at large $N$ is from 
configurations where the first $\nu_+N$ eigenvalues are around one of the minima 
($\mu$) and the remaining $\nu_-N$ are around the other ($-\mu$). 
Note that flipping the signs of the $\nu_-N$ eigenvalues: $\lambda_j \to -\lambda_j$ ($j=\nu_+N+1,\cdots,N$) 
in (\ref{Zff}) leads to 
\be
Z_{(\nu_+,\nu_-)}=(-1)^{\nu_-N}Z_{(1,0)}. 
\label{Zff_Z10}  
\ee
Consequently, the total partition function vanishes~\footnote{
As discussed in~\cite{Kuroki:2010au}, the total partition function can be regarded as a zero-dimensional analog of 
the Witten index~\cite{Witten:1982df}.}: 
\be
Z = \sum_{\nu_-N=0}^N\frac{N!}{(\nu_+N)!(\nu_-N)!}\,Z_{(\nu_+,\nu_-)} = (1+(-1))^N\,Z_{(1,0)}=0, 
\label{Z_Zff}
\ee
rendering expectation values normalized by the partition function ill-defined or indefinite. 
Here, we regularize the partition function by introducing a factor $e^{-i\alpha\nu_-N}$ with small $\alpha$ in front of 
$Z_{(\nu_+,\nu_-)}$. This corresponds to assigning the phase $e^{-i\alpha}$ 
to each integration measure over the negative real axis $d\lambda_j$ ($j=\nu_+N+1,\cdots,N$) 
in (\ref{Zff}). 
The regularized partition function becomes 
\be
Z_\alpha\equiv\sum_{\nu_-N=0}^N\frac{N!}{(\nu_+N)!(\nu_-N)!}\,e^{-i\alpha\nu_-N}Z_{(\nu_+,\nu_-)} 
= (1-e^{-i\alpha})^N\,Z_{(1,0)}. 
\label{Z_alpha}
\ee
The phase $\alpha$ is reminiscent of an external field discussed in~\cite{Kuroki:2009yg,Kuroki:2010au},
which was introduced in order to observe whether the supersymmetry is spontaneously broken or not. 

Since the auxiliary field $B$ in (\ref{S}) is invariant under the supersymmetry transformations 
generated by $Q$ and $\bar{Q}$, 
the expectation value 
$
\vev{\frac{1}{N}\tr\,(iB)}_\alpha =\vev{\frac{1}{N}\tr\,(\phi^2-\mu^2)}_{\alpha} 
$
taken with respect to the regularized partition function (\ref{Z_alpha}) 
will play the role of an order parameter for spontaneous breaking of the supersymmetry, 
provided the limit $\alpha\to 0$ is well-defined. 
Noting~\footnote{
The superscript $(1,0)$ on the left hand side (l.h.s.) of (\ref{vevB_Z10}) 
indicates an expectation value taken with respect to the partition function $Z_{(1,0)}$.} 
\be
\vev{\frac{1}{N}\tr\,(\phi^2-\mu^2)}^{(1,0)}=\frac{1}{N^2}\,\frac{1}{Z_{(1,0)}}\,
\frac{\partial}{\partial(\mu^2)}Z_{(1,0)}
\label{vevB_Z10}
\ee
from (\ref{Zff}) with $(\nu_+, \nu_-)=(1,0)$, we see that $\vev{\frac{1}{N}\tr\,(iB)}_\alpha$ coincides with 
$\vev{\frac{1}{N}\tr\,(\phi^2-\mu^2)}^{(1,0)}$:
\be
\vev{\frac{1}{N}\tr\,(iB)}_\alpha=
\vev{\frac{1}{N}\tr\,(\phi^2-\mu^2)}_{\alpha} \equiv \frac{1}{N^2}\,\frac{1}{Z_{\alpha}}\, 
\frac{\partial}{\partial(\mu^2)}Z_{\alpha} = \vev{\frac{1}{N}\tr\,(\phi^2-\mu^2)}^{(1,0)} 
\label{vevB_Z_alpha}
\ee
due to a cancellation of the factor $(1-e^{-i\alpha})^N$ in (\ref{Z_alpha}) between the numerator and the denominator. 
The regularized expectation value $\vev{\frac{1}{N}\tr\,(iB)}_{\alpha}$ is 
independent of $\alpha$ and well-defined in the limit $\alpha\to 0$, and thus serves as an order parameter. 

Perturbative contributions to $Z_{(1,0)}$ are computed by a deformation method used 
in topological field theory in appendix~\ref{app:localization}, and the result is 
\be
\left.Z_{(1,0)}\right|_{{\rm pert.}} =1  
\label{Z10_localization}
\ee 
for arbitrary $N$. 
Notice that (\ref{Z10_localization}) 
is valid to all orders in the perturbation around the saddle point $\lambda_i=\mu$ ($i=1,\cdots, N$), 
{\it but excludes nonperturbative effects}. 
Combining this result with (\ref{vevB_Z10}) and (\ref{vevB_Z_alpha}) suggests that the supersymmetry 
is unbroken within perturbation theory. 
In the following, we consider nonperturbative effects on quantities in a double scaling limit 
that realizes a nonperturbative formulation of the corresponding string theory.  
As discussed in~\cite{Kuroki:2012nt,Kostov:1990wi,Kostov:1992pn,Gaiotto:2004nz}, perturbative contributions to 
correlation functions among operators 
of even powers of $\phi$ are described by the $c=-2$ topological gravity where the string susceptibility 
exponent is $\gamma=-1$. Thus, for the double scaling limit, we consider the case of $\mu^2$ approaching the critical point as $\mu^2\to 2+0$, 
(i.e. $\omega\to +0$ in $\mu^2=2+4\omega$)~\cite{Kuroki:2012nt} while sending $N$ to infinity 
such that the combination $N^2\omega^{2-\gamma}=N^2\omega^3$ is fixed. 
Assuming that this limit is also valid for nonperturbative effects, we take the scaling variable 
\be
t\equiv N^{2/3}\omega 
\label{dsl}
\ee
to be fixed in the double scaling limit. 
According to the correspondence discussed in~\cite{Kuroki:2012nt,Kuroki:2013qpa}, 
the double scaling limit is expected to give a nonperturbative framework 
for two-dimensional type IIA superstring theory 
on a nontrivial Ramond-Ramond background, where $t^{-3/2}$ plays the role of a renormalized string coupling 
constant 
and the strength of the background is related to $(\nu_+-\nu_-)$.

Although for any finite $N$ the supersymmetry is spontaneously broken by a tunneling (instanton) effect between 
the minima $\pm \mu$ of the double-well, 
the effect ceases in a simple large-$N$ limit ($N\to \infty$ with $\omega$ fixed) and the supersymmetry 
becomes restored~\cite{Witten:1982df,Affleck:1983pn}. 
However, we should notice that it is a nontrivial question how the situation goes 
in the double scaling limit. In fact, we will see in the following that the supersymmetry breaking remains 
after the double scaling limit. 
Due to the correspondence~\cite{Kuroki:2013qpa}, 
nonperturbative dynamics in the two-dimensional type IIA superstring is expected to induce supersymmetry 
breaking in the target space.

\section{Instanton effects in the matrix model}
\label{sec:MMinst}
\setcounter{equation}{0}
In this section, we consider effects of instantons in the matrix model by a method similar to what is discussed 
in~\cite{Hanada:2004im,Ishibashi:2005dh}. 

The partition function in the $(1,0)$ sector $Z_{(1,0)}$ is given by integrals along the positive real axis with respect to all 
$N$ eigenvalues. Its perturbative contribution (contribution without instanton effects) 
at large $N$ comes from the integration region 
$[a, b]$ with 
\be
a=\sqrt{\mu^2-2}, \qquad b=\sqrt{\mu^2+2},
\ee
which is nothing but the support of the eigenvalue distribution~\footnote{
The eigenvalue distribution for a general filling fraction $(\nu_+, \nu_-)$ is 
\be
\rho_{(\nu_+,\nu_-)}(x) = \left\{\begin{array}{cl} \frac{\nu_+}{\pi}\,x\,\sqrt{(x^2-a^2)(b^2-x^2)} & \qquad (a<x<b) \\
                                   \frac{\nu_-}{\pi}\,|x|\,\sqrt{(x^2-a^2)(b^2-x^2)} & \qquad (-b<x<-a).
                 \end{array}\right. 
\label{rho_nupm}                 
\ee 
(\ref{rho_nupm}) and (\ref{eigenvalue_distribution}) are obtained in sections 4.1.1 and 4.1.2 
of ref.~\cite{Kuroki:2009yg}. 
Note that the notation $\mu^2$ there corresponds to $-\mu^2$ here. 
}
\be
\rho_{(1,0)}(x)\equiv \vev{\frac{1}{N}\sum_{i=1}^N\delta(x-\lambda_i)}^{(1,0)}_{\rm planar}
=\frac{x}{\pi}\sqrt{(x^2-a^2)(b^2-x^2)}. 
\label{eigenvalue_distribution}
\ee 
The suffix ``planar'' associated with the expectation value means to take planar contributions.
We divide the region ${\bf R}_+\equiv [0,\infty)$ for each eigenvalue into the support and its complement, 
and express the partition function as 
\be
Z_{(1,0)}= \sum_{k=0}^N \left.Z_{(1,0)}\right|_{k{\rm -inst.}}.
\ee
The term involving $k$ eigenvalues integrated over the outside of the support is regarded as  
the $k$-instanton contribution $\left.Z_{(1,0)}\right|_{k{\rm -inst.}}$, and is given by 
\bea
\left. Z_{(1,0)}\right|_{k{\rm -inst.}} & \equiv & \begin{pmatrix} N \\ k \end{pmatrix} \tilde{C}_N \int_a^b\prod_{i=1}^{N-k}d\lambda_i 
\int_{{\bf R}_+-[a,b]}\prod_{j=N-k+1}^N d\lambda_j \left(\prod_{n=1}^N 2\lambda_n\right) 
\left\{\prod_{n>m}(\lambda_n^2-\lambda_m^2)^2\right\} \nn \\
& & \hspace{56mm} \times e^{-N\sum_{i=1}^N\frac12(\lambda_i^2-\mu^2)^2} 
\eea
in accordance with~\cite{Hanada:2004im}. Since the result (\ref{Z10_localization}) implies 
\be
\left.Z_{(1,0)}\right|_{0{\rm -inst.}}=1,
\label{Z0inst}
\ee
the free energy $F_{(1,0)}\equiv -\ln Z_{(1,0)}$ can be expressed as 
\bea
F_{(1,0)} & = & -\ln\left[ \left.Z_{(1,0)}\right|_{0{\rm -inst.}}+\left.Z_{(1,0)}\right|_{1{\rm -inst.}}+\cdots\right] \nn \\
& = & - \left.Z_{(1,0)}\right|_{1{\rm -inst.}}+\cdots, 
\label{Fenergy}
\eea
where the omitted terms represent contributions from multi-instantons. 

In order to evaluate the one-instanton contribution, we choose $y\equiv \lambda_N$ and rewrite  
$\left.Z_{(1,0)}\right|_{1{\rm -inst.}}$ as 
\be
\left.Z_{(1,0)}\right|_{1{\rm -inst.}} = N\hspace{-1mm}
\left.Z_{(1,0)}'\right|_{0{\rm -inst.}}\int_{{\bf R}_+-[a,b]} 
2ydy\,e^{-\frac{N}{2}(y^2-\mu^2)^2}\,\vev{\prod_{i=1}^{N-1}(y^2-\lambda_i^2)^2}^{'\,(1,0)}. 
\label{Z1inst_y}
\ee
Here, quantities with a prime ($\,'\,$) concern the system of $N-1$ eigenvalues $\lambda_i$ ($i=1,\cdots, N-1$). 
Explicitly, 
\be
\left.Z_{(1,0)}'\right|_{0{\rm -inst.}}\equiv \tilde{C}_N\int_a^b\prod_{i=1}^{N-1}(2\lambda_id\lambda_i)\,
\left\{\prod_{N-1\geq i>j\geq 1}(\lambda_i^2-\lambda_j^2)^2\right\}\,
e^{-N\sum_{i=1}^{N-1}\frac12(\lambda_i^2-\mu^2)^2}, 
\ee
and the expectation value 
$\vev{\prod_{i=1}^{N-1}(y^2-\lambda_i^2)^2}^{'\,(1,0)}$ 
is taken with respect to the partition function $\left.Z_{(1,0)}'\right|_{0{\rm -inst.}}$. 
The expectation value is expanded in cumulants:  
\bea
& & \vev{\prod_{i=1}^{N-1}(y^2-\lambda_i^2)^2}^{'\,(1,0)}
= \vev{e^{2{\rm Re}\,\sum_{i=1}^{N-1}\ln(y^2-\lambda_i^2)}}^{'\,(1,0)} \nn \\
& & = \exp\Bigg[\vev{2{\rm Re}\,\sum_{i=1}^{N-1}\ln(y^2-\lambda_i^2)}^{'\,(1,0)} 
+\frac12\vev{\left\{2{\rm Re}\,\sum_{i=1}^{N-1}\ln(y^2-\lambda_i^2)\right\}^2}_C^{'\,(1,0)}+\cdots\Bigg] \nn \\
& & = \exp\Bigg[\vev{2{\rm Re}\,\sum_{i=1}^N\ln(y^2-\lambda_i^2)}_{\rm planar}^{(1,0)}+\Delta_0D(y^2)
+\frac12\vev{\left\{2{\rm Re}\,\sum_{i=1}^N\ln(y^2-\lambda_i^2)\right\}^2}_{C, \,{\rm planar}}^{\,(1,0)} \nn \\
 & & \hspace{14mm}+ \cO(N^{-1})\Bigg] \ , 
\label{cumulant}
\eea
where the suffix $C$ indicates taking the connected parts, 
and the leading order contribution to the exponent is given by the disk amplitude 
$\vev{2{\rm Re}\,\sum_{i=1}^N\ln(y^2-\lambda_i^2)}_{\rm planar}^{(1,0)}$, which is of order $N$. 
For now, we count the order of $N$ in a simple manner (with $\omega$ fixed).
Contributions to the exponent at subleading order $\cO(N^0)$ consist of the difference in 
disk amplitudes 
\bea
\Delta D(y^2) & \equiv & \vev{2{\rm Re}\,\sum_{i=1}^{N-1}\ln(y^2-\lambda_i^2)}_{\rm planar}^{'\,(1,0)}-\vev{2{\rm Re}\,\sum_{i=1}^N\ln(y^2-\lambda_i^2)}_{\rm planar}^{(1,0)} \nn \\
& = & \Delta_{0}D(y^2) +\Delta_{1}D(y^2)+\cdots
\label{DeltaDy2}
\eea
($\Delta_nD(y^2)$ denotes the $\cO(N^{-n})$ part of the difference $\Delta D(y^2)$) and 
the annulus amplitude 
$\frac12\vev{\left\{2{\rm Re}\,\sum_{i=1}^N\ln(y^2-\lambda_i^2)\right\}^2}_{C,\,{\rm planar}}^{\,(1,0)}$. 
The $\cO(N^{-1})$ terms in the exponent in (\ref{cumulant}) comes from higher-point 
or higher-genus amplitudes 
of the loop operator $2{\rm Re}\,\sum_{i=1}^{N-1}\ln(y^2-\lambda_i^2)$ and from the difference in 
annulus amplitudes defined similar to $\Delta D(y^2)$. 
Here, we take into account contributions up to $\cO(N^0)$ 
as discussed in~\cite{Hanada:2004im,Ishibashi:2005dh}. 
Then, (\ref{Z1inst_y}) is expressed as
\be
\left.Z_{(1,0)}\right|_{1{\rm -inst.}}=N\hspace{-1mm}\left.Z_{(1,0)}'\right|_{0{\rm -inst.}}
\int_{{\bf R}_+-[a,b]} 2ydy\,e^{-N V_{\rm eff}^{(0)}(y)-V_{\rm eff}^{(1)}(y)+\cO(N^{-1})} 
\label{Z1inst_y2}
\ee
with 
\bea
V_{\rm eff}^{(0)}(y) & \equiv & \frac12(y^2-\mu^2)^2-2{\rm Re}\,\vev{\frac1N\sum_{i=1}^N\ln(y^2-\lambda_i^2)}_{\rm planar}^{(1,0)}, 
\label{Veff0} \\
V_{\rm eff}^{(1)}(y) & \equiv & -\Delta_0D(y^2) -\frac12\vev{\left\{2{\rm Re}\,\sum_{i=1}^N\ln(y^2-\lambda_i^2)\right\}^2}_{C,\,{\rm planar}}^{\,(1,0)}.
\label{Veff1}
\eea  
$V_{\rm eff}^{(0)}(y)$ and $V_{\rm eff}^{(1)}(y)$ represent the potential felt by the eigenvalue $y$ at 
leading and subleading orders in $1/N$, respectively. 

\begin{figure}
\centering
\includegraphics[height=7cm, width=12cm, clip]{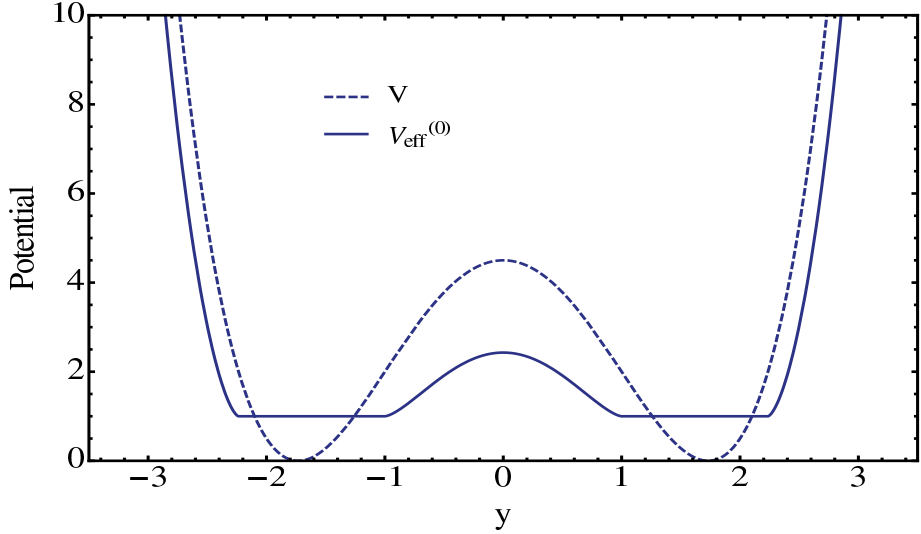}
\caption{The dashed and solid curves show the double-well potential $V(y)=\frac12(y^2-\mu^2)^2$ 
and the effective potential $V_{\rm eff}^{(0)}(y)$ respectively for $\mu^2=3$. } 
\label{fig:Veff0}
\end{figure}

\paragraph{Contribution from $V_{\rm eff}^{(0)}$} 
The planar expectation value of the resolvent is computed in appendix A of ref.~\cite{Kuroki:2012nt}
and is given by 
\be
\vev{\frac1N\sum_{i=1}^N\frac{1}{z-\lambda_i^2}}^{(1,0)}_{\rm planar}=\frac12\left[z-\mu^2-\sqrt{(z-a^2)(z-b^2)}\right]. 
\label{R2z}
\ee 
The second term in (\ref{Veff0}) is obtained by integrating (\ref{R2z}) with respect to $z$:
\be
\vev{\frac{1}{N}\sum_{i=1}^N\ln(z-\lambda_i^2)}^{(1,0)}_{\rm planar}=\lim_{\Lambda\to\infty}\left[
\int^z_\Lambda dz'\vev{\frac{1}{N}\sum_{i=1}^N\frac{1}{z'-\lambda_i^2}}^{(1,0)}_{\rm planar}+\ln \Lambda\right]. 
\label{trln_disk}
\ee
It is derived from a comparison of the large-$z$ expansions of $\ln(z-\lambda_i^2)$ and 
$\frac{1}{z-\lambda_i^2}$.   
Then, the effective potential $V_{\rm eff}^{(0)}(y)$ becomes  
\be
V_{\rm eff}^{(0)}(y)= V_{\rm eff}^{(0)}(b)+\begin{cases} \int^{y^2}_{b^2}dz\,\sqrt{(z-a^2)(z-b^2)} & (|y|>b) \\
                             \int^{a^2}_{y^2}dz\,\sqrt{(a^2-z)(b^2-z)} & (|y|<a) \\
                               0  &  (a<|y|<b).
                      \end{cases}         
\label{Veff0_2}                      
\ee
Although it is sufficient to consider the case of real positive $y$ for the filling fraction $(1,0)$, 
the expression of $V_{\rm eff}^{(0)}(y)$ can be naturally extended to negative $y$. 
Note that the r.h.s. of (\ref{R2z}) and thus (\ref{Veff0_2}) are valid for a general filling fraction. 
The potential is flat and the eigenvalue $y$ feels no force 
within the support of the eigenvalue distribution $[a,b]$ (or $[-b,-a]$). 
This can be understood from the fact that $y$ receives no net force in the sea of the other 
eigenvalues~\cite{Hanada:2004im}. 
After calculating the integrals, we find
\be
V_{\rm eff}^{(0)}(y)=1+\frac12\,|y^2-\mu^2|\sqrt{(y^2-\mu^2)^2-4}-2\ln\frac{|y^2-\mu^2|+\sqrt{(y^2-\mu^2)^2-4}}{2}
\ee 
for $|y|>b$ or $|y|<a$. In particular, we obtain $V_{\rm eff}^{(0)}(b)=1$. 
The form of $V_{\rm eff}^{(0)}(y)$ is plotted in Fig.~\ref{fig:Veff0}. 
It has a local maximum at $y=0$, whose value is  
\be
V_{\rm eff}^{(0)}(0)=1+\frac12\,\mu^2\sqrt{\mu^4-4}-2\ln\frac{\mu^2+\sqrt{\mu^4-4}}{2}
=1+\frac{32}{3}\,\omega^{3/2} + \cO(\omega^{5/2}).
\label{Veff0at0}
\ee
%
%

For the $y$-integration in (\ref{Z1inst_y2}), we focus on a region near the origin that would be responsible for
contributions from instantons, as discussed in~\cite{Hanada:2004im}. 
For $|y|<a$, we find: 
\be
N\left(V_{\rm eff}^{(0)}(y)-1\right) = N\left[\frac43\,(a^2-y^2)^{3/2}+\frac{1}{10}\,(a^2-y^2)^{5/2}+ \cO\left((a^2-y^2)^{7/2}\right)\right], 
\label{NVeff_I}
\ee
where only the first term survives in the double scaling limit, and the integration near the origin  
leads to 
\bea
\int_0 2ydy\,e^{-N(V^{(0)}_{\rm eff}(y)-1)} & = & \int^{a^2}ds\,e^{-\frac43\,Ns^{3/2}}=\frac{4t}{N^{2/3}}\int^1ds\,e^{-\frac{32}{3}\,t^{3/2}s^{3/2}} 
\nn \\
 & = & -\frac{1}{4N^{2/3}t^{1/2}}\,e^{-\frac{32}{3}\,t^{3/2}}\,\left(1+\cO(t^{-3/2})\right)
\label{int_exp_Veff0} 
\eea
for the case of $t$ finite but large. 
The exponent is nothing but the height of the potential barrier of $NV^{(0)}_{\rm eff}(y)$: 
\be
e^{-N(V_{\rm eff}^{(0)}(0)-1)}=e^{-\frac{32}{3}\,t^{3/2}}. 
\label{expVeff0}
\ee
For the contribution from the potential at subleading order $V_{\rm eff}^{(1)}(y)$, we substitute it with 
the value at the origin $V_{\rm eff}^{(1)}(0)$.  
Then, the free energy (\ref{Fenergy}) becomes 
\be
F_{(1,0)} = Ne^{-N}\hspace{-1mm}
\left.Z'_{(1,0)}\right|_{0{\rm -inst.}}\frac{e^{-V_{\rm eff}^{(1)}(0)}}{4N^{2/3}t^{1/2}}\,
e^{-\frac{32}{3}\,t^{3/2}}\,\left(1+\cO(t^{-3/2})\right) + \cdots. 
\label{Fenergy_2}
\ee 

\paragraph{Contribution from the remaining factors} 
Next, let us evaluate the contribution from the factor $\left.Z'_{(1,0)}\right|_{0{\rm -inst.}}$ in 
(\ref{Fenergy_2}). 
Taking
\be
\lambda_i=\left(\frac{N-1}{N}\right)^\frac14\lambda_i', 
\quad \mu=\left(\frac{N-1}{N}\right)^\frac14\mu', \quad 
a=\left(\frac{N-1}{N}\right)^\frac14 a', \quad b=\left(\frac{N-1}{N}\right)^\frac14 b'\ ,
\label{replacement_Z'}
\ee
one finds 
\bea
\left.Z'_{(1,0)}\right|_{0{\rm -inst.}} & = & \frac{\tilde{C}_N}{\tilde{C}_{N-1}}
\left(\frac{N-1}{N}\right)^{\frac{(N-1)^2}{2}} \nn \\
& & \times \tilde{C}_{N-1}\int^{b'}_{a'}\prod_{i=1}^{N-1}(2\lambda_i'd\lambda_i')
\left\{\prod_{N-1\geq i>j\geq 1}(\lambda_i'^2-\lambda_j'^2)^2\right\} 
e^{-(N-1)\sum_{i=1}^{N-1}\frac12(\lambda_i'^2-\mu'^2)^2}. \nn \\
& & \label{Z'0inst2}
\eea
Note that the last line in (\ref{Z'0inst2}) is nothing but $\left.Z_{(1,0)}\right|_{0{\rm -inst.}}$ 
with the replacements 
\be
N\to N-1, \qquad \mu\to \mu',\qquad a\to a', \qquad b\to b'. 
\ee
Therefore, the last line in (\ref{Z'0inst2}) equals unity by a perturbative argument around the saddle point 
$\lambda_i'=\mu'$ ($i=1,\cdots, N$) 
which is parallel to the derivation of (\ref{Z0inst}), i.e. (\ref{Z10_pertf}). 
By using (\ref{CtildeN}), we obtain 
\be
\left.Z'_{(1,0)}\right|_{0{\rm -inst.}}=\frac{\tilde{C}_N}{\tilde{C}_{N-1}}
\left(\frac{N-1}{N}\right)^{\frac{(N-1)^2}{2}}=\frac{e^N}{2\pi N}\times \left(1+\cO(N^{-1})\right).
\label{Z'0inst_f}
\ee 
Also, $V_{\rm eff}^{(1)}(0)$ is computed in appendix~\ref{app:Veff1}, 
and the result (\ref{app:Veff1_f}) gives  
\be
e^{-V_{\rm eff}^{(1)}(0)}=\frac{N^{2/3}}{16t}\times \left(1+\cO(N^{-1/3})\right). 
\label{exp_Veff1}
\ee

\paragraph{Final result} 
Plugging (\ref{Z'0inst_f}) and (\ref{exp_Veff1}) into (\ref{Fenergy_2}), we see that the double scaling limit 
leaves a finite and nontrivial function of $t$: 
\be
F_{(1,0)}= \frac{1}{128\pi\,t^{3/2}}\,e^{-\frac{32}{3}t^{3/2}}\,\left(1+\cO(t^{-3/2})\right) + 
(\mbox{multi-instantons}) 
\label{Fenergy_f}
\ee 
for $t$ finite but large. The result supports the validity of taking (\ref{dsl}) as a scaling variable. 
Similar to the $c=0$ case~\cite{Hanada:2004im}, it would be natural to regard instantons in the matrix model 
as kinds of D-branes in the corresponding type IIA superstring theory
in two dimensions~\cite{Kuroki:2012nt,Kuroki:2013qpa}. 
In fact, $e^{-\frac{32}{3}t^{3/2}}$ is essentially the exponential series of the disk amplitude 
whose boundary is placed at the position of the instanton $y=0$: 
$\left.\vev{2{\rm Re}\,\sum_{i=1}^N\ln(y^2-\lambda_i^2)}_{\rm planar}^{(1,0)}\right|_{y=0}$, 
which seems parallel to the argument in~\cite{Polchinski:1994fq}.  
The remaining factor $\frac{1}{128\pi\,t^{3/2}}$ in (\ref{Fenergy_f}) 
receives contributions 
from fluctuation in the position of the instanton and from the exponential of the annulus amplitude at the 
origin: 
$
\left.\frac12\vev{\left\{2{\rm Re}\,\sum_{i=1}^N\ln(y^2-\lambda_i^2)\right\}^2}_{C,\,{\rm planar}}^{\,(1,0)}\right|_{y=0}
$. 
The difference of disk amplitudes $\Delta_0D(0)$ does not contribute in the double scaling limit as seen 
in appendix~\ref{app:Veff1}. 
This would be clarified by considering analogs of FZZT or ZZ 
branes~\cite{Fateev:2000ik,Teschner:2000md,Zamolodchikov:2001ah} 
in the type IIA superstring theory and computing 
amplitudes in the presence of such branes. We leave it as a future subject for investigation.  

We also comment on two notable points which differ from the situation for the $c=0$ case. 
First, the instanton effect (\ref{Fenergy_f}) is a real number, 
while it is pure imaginary in the $c=0$ case~\cite{Hanada:2004im} indicating instability of that 
system. Technically, the result of the latter is attributed to rotating the integration path of an eigenvalue 
in order to obtain a finite result. 
Our computation does not need such a rotation of the integration path, 
and the result does not seem to exhibit any 
instability~\footnote{
In a double-well matrix model consisting only of the bosonic part of our matrix model,  
instanton effects are computed by rotating the integration path~\cite{Kawai:2004pj} similar to the $c=0$ case, 
and the result is imaginary valued. Since that case seems to be well-defined without rotating the path, 
it is expected to yield real and finite instanton effects by integrating along the original contour. 
}.  
This provides evidence that our matrix model leads to a sensible theory in the double scaling limit. 
Second, the powers of $N$ appearing in (\ref{Fenergy_f}) are integers by recalling (\ref{dsl}), while  
in the $c=0$ case~\cite{Hanada:2004im} they are half-integers~\footnote{
According to refs.~\cite{Alexandrov:2003nn,Ishibashi:2005zf}, 
this is also the case for minimal string theories.}. 
Our result tempts us to interpret contributions to (\ref{Fenergy_f}) as string worldsheets with holes 
at the positions of instantons based on the identification of $1/N$ as a string coupling. 
However, such an interpretation does not seem
 straightforward when considering contributions from 
the $y$-integral which represent the fluctuations in the position of the instanton.

\section{Orthogonal polynomials}
\label{sec:ortho_poly}
\setcounter{equation}{0}
In sections~\ref{sec:1-inst} and {\ref{sec:2-inst}, we compute nonperturbative effects including 
the result obtained in the previous section in a more efficient way. 
In preparation for this, let us first introduce orthogonal polynomials for the matrix model in this section. 

Under the change of variables $x_i=\lambda_i^2-\mu^2$, the partition function $Z_{(1,0)}$ defined 
in (\ref{Zff}) 
reduces to Gaussian matrix integrals  
\be
Z_{(1,0)} = \tilde{C}_N\int^\infty_{-\mu^2}\left(\prod_{i=1}^N dx_i\right)\,\triangle(x)^2\,
e^{-N\sum_{i=1}^N \frac12 x_i^2}. 
\label{Z10_x}
\ee
It seems almost trivial, but a nontrivial effect possibly arises from the boundary of the integration region. 
Ref.~\cite{Gaiotto:2004nz} mentions that the boundary effect is nonperturbative in $1/N$. 
Indeed, if we neglect it by replacing the lower bound $-\mu^2$ with $-\infty$, (\ref{Z10_x}) will coincide 
with the perturbative result (\ref{Z10_localization}) or (\ref{Z0inst}). 
This suggests that the supersymmetry is unbroken to all orders in the $1/N$ expansion. 

Let us consider polynomials 
\be
P_n(x)=x^n+\sum_{i=0}^{n-1}p_n^{(i)}x^i\qquad (n=0,1,2,\cdots)
\ee
with the coefficient of the top degree ($x^n$) fixed to 1. The coefficients $p_n^{(i)}$ are uniquely 
determined so that the orthogonality relation 
\be
(P_n,\,P_m)\equiv \int^\infty_{-\mu^2}dx\,e^{-\frac{N}{2}x^2}\,P_n(x)\,P_m(x) = h_n\delta_{n,m}
\label{orthogonal}
\ee
is satisfied. Similar to the case without a boundary~\cite{Itzykson:1979fi}, we have recursion 
relations of the form 
\bea
& & xP_m(x)=P_{m+1}(x)+S_m P_m(x)+R_mP_{m-1}(x), 
\label{recursion_P}\\
& & h_m=R_mh_{m-1}.
\label{hn_Rn}
\eea
For example, the first few quantities are 
\bea
& & h_0 = \sqrt{\frac{2\pi}{N}}\left[1-\frac12\,\mbox{erfc}\left(\sqrt{\frac{N}{2}}\,\mu^2\right)\right], 
\label{h0} \\
& & S_0=-p_1^{(0)}=\frac{1}{Nh_0}\,e^{-\frac{N}{2}\mu^4}, \label{S0}\\
& & h_1=\frac{1}{N}\,h_0-\frac{1}{N}\,\mu^2\,e^{-\frac{N}{2}\,\mu^4}-\frac{1}{N^2h_0}\,e^{-N\mu^4},  
\eea
where 
\be
\mbox{erfc}(x)\equiv \frac{2}{\sqrt{\pi}}\int^\infty_x dt\,e^{-t^2}.
\ee
Then, (\ref{Z10_x}) and the expectation value $\vev{\frac{1}{N}\tr\,(\phi^2-\mu^2)}^{(1,0)}$ are expressed as 
\be
Z_{(1,0)} = \tilde{C}_N\,N!\left(\prod_{n=0}^{N-1}h_n\right) 
\label{Z10_h}
\ee
and 
\bea
\vev{\frac{1}{N}\tr\,(\phi^2-\mu^2)}^{(1,0)} & = & 
\frac{1}{Z_{(1,0)}}\,\tilde{C}_N\int^\infty_{-\mu^2}\left(\prod_{i=1}^N dx_i\right)\,\triangle(x)^2\,
e^{-N\sum_{i=1}^N \frac12 x_i^2}\,\frac{1}{N}\sum_{k=1}^N x_k \nn\\
& = &  \frac{1}{N}\sum_{n=0}^{N-1}S_n, 
\label{vevB_S}
\eea
respectively. 
We will compute (\ref{vevB_S}) later by taking into account the boundary effect.
The identities 
\bea
& & \int^{\infty}_{-\mu^2} dx\,\frac{d}{dx}
\left(e^{-\frac{N}{2}x^2}\,P_n(x)^2\right)=-e^{-\frac{N}{2}\mu^4}\,P_n(-\mu^2)^2,  \\
& & \int^{\infty}_{-\mu^2} dx\,\frac{d}{dx}
\left(e^{-\frac{N}{2}x^2}\,P_n(x)P_{n-1}(x)\right)=-e^{-\frac{N}{2}\mu^4}\,P_n(-\mu^2)P_{n-1}(-\mu^2)
\eea 
lead to relations which include the boundary effects:
\bea
S_n & = & \frac{1}{N}\frac{1}{h_n}\,P_n(-\mu^2)^2\,e^{-\frac{N}{2}\mu^4}, 
\label{Sn}\\
R_n & = & \frac{n}{N}+\frac{1}{N}\frac{1}{h_{n-1}}\,P_n(-\mu^2)P_{n-1}(-\mu^2)\,e^{-\frac{N}{2}\mu^4}.  
\label{Rn}
\eea

As an approximation of the zeroth order contributions, let us simply neglect the boundary effect. 
This corresponds to changing the lower bound of the integral (\ref{orthogonal}) to $-\infty$ 
and dropping terms containing $e^{-\frac{N}{2}\mu^4}$ in (\ref{Sn}) and (\ref{Rn}).  
The orthogonal polynomials in this case, denoted by $P^{(H)}_n(x)$, are given by the Hermite polynomials:
\be
P^{(H)}_n(x)=\frac{1}{(2N)^{n/2}}\,H_n\left(\sqrt{\frac{N}{2}}\,x\right)
\label{PH}
\ee
with 
\be
H_n(x) \equiv (-1)^n\,e^{x^2}\frac{d^n}{dx^n}\,e^{-x^2} 
\ee
and coefficients 
\be
S_n^{(H)}=0,\qquad R^{(H)}_n=\frac{n}{N}, \qquad h_n^{(H)}=\sqrt{2\pi}\,\frac{n!}{N^{n+\frac12}}.
\label{SRh_Hn}
\ee
The superscript $(H)$ represents quantities in the zeroth order approximation. 

We will compute corrections due to the boundary, denoted by quantities with tildes: 
\bea
& & P_n(x)=P^{(H)}_n(x)+\tilde{P}_n(x), \nn \\
& & S_n=S^{(H)}_n+\tilde{S}_n, \qquad R_n=R^{(H)}_n +\tilde{R}_n, \qquad h_n=h^{(H)}_n+\tilde{h}_n 
\label{PSRh_H_tilde}
\eea 
in an iterative manner. 
In terms of the ratio 
\be
k_m(x)=\frac{P_m(x)}{P_{m-1}(x)}, 
\ee
(\ref{recursion_P}) is expressed as 
\be
x=k_{m+1}(x)+S_m+\frac{R_m}{k_m(x)}. 
\label{recursion_k}
\ee
The zeroth and first order contributions to (\ref{recursion_k}) with respect to the corrections are 
\be
 x=k^{(H)}_{m+1}(x)+\frac{m}{N}\,\frac{1}{k_m^{(H)}(x)} 
\label{recursion_0th_k}  
\ee
and
\be
  0 = k_{m+1}^{(H)}(x) \tilde{L}_{m+1}(x) -x\tilde{L}_m(x) +\frac{1}{k_m^{(H)}(x)}\frac{m}{N}\tilde{L}_{m-1}(x) 
 +\tilde{S}_m+\frac{1}{k_m^{(H)}(x)}\tilde{R}_m,
 \label{recursion_1st_k}
\ee
where 
\be
k^{(H)}_m(x)\equiv \frac{P^{(H)}_m(x)}{P^{(H)}_{m-1}(x)}, \qquad 
\tilde{L}_m(x)\equiv \frac{\tilde{P}_m(x)}{P_m^{(H)}(x)}. 
\ee
We expand quantities with tildes in terms of instanton number as:
\bea
\tilde{S}_n & = & \tilde{S}^{(1)}_n + \tilde{S}^{(2)}_n + \cdots, \nn \\
\tilde{R}_n & = & \tilde{R}^{(1)}_n + \tilde{R}^{(2)}_n + \cdots, \nn \\
\tilde{h}_n & = & \tilde{h}^{(1)}_n + \tilde{h}^{(2)}_n +\cdots, \nn \\
\tilde{L}_n(x) & =& \tilde{L}^{(1)}_n(x) + \tilde{L}^{(2)}_n(x) +\cdots. 
\eea
The superscripts $(1), (2), \cdots$ represent contributions from one instanton, two instantons and so forth.

\section{One-instanton contribution}
\label{sec:1-inst}
\setcounter{equation}{0}
In this section, we consider (\ref{recursion_0th_k}) as the first iteration 
with respect to the instanton number. 
The obtained result is interpreted as a nonperturbative effect from a one-instanton configuration. 

\subsection{Leading order} 
\label{sec:1-inst_leading}
Following the argument in section~3.2 of~\cite{Hanada:2004im}, 
we assume that $k^{(H)}_m(x)$ has smooth large-$N$ 
behavior given by
\be
k^{(H)}_m(x) =k^{(0)}(x,\xi)+\frac{1}{N}\,k^{(1)}(x, \xi)+\cO(N^{-2})
\label{kH}
\ee
with $\xi=\frac{m}{N}$. Then, $\cO(N^0)$ and $\cO(N^{-1})$ contributions to (\ref{recursion_0th_k}) 
determine $k^{(0)}(x, \xi)$ and $k^{(1)}(x,\xi)$ to be
\bea
k^{(0)}(x,\xi) & = & \mbox{sgn}(x)\frac{|x|+\sqrt{x^2-4\xi}}{2} \qquad \mbox{for}\qquad |x|>2, \nn \\
k^{(1)}(x,\xi) & = & -\frac12k^{(0)}(x,\xi)\partial_\xi \ln \sqrt{x^2-4\xi}.
\label{k0k1}
\eea
The orthogonal polynomial $P^{(H)}_n(x)$ for $|x|>2$ can be expressed by 
\bea
P^{(H)}_n(x) & = & \prod_{m=1}^n k^{(H)}_m(x) \nn \\
 & = & (\mbox{sgn}(x))^n\,\exp\left[\sum_{m=1}^n \ln|k^{(0)}(x, \xi)| 
+\frac{1}{N}\sum_{m=1}^n \frac{k^{(1)}(x, \xi)}{k^{(0)}(x, \xi)}+\cO(N^{-1})\right],
\label{PHn}
\eea
where we consider $n$ running up to $N-1$. 
The Euler-Maclaurin formula 
\be
\sum_{m=1}^n f\left(\frac{m}{N}\right)=N\int^{\frac{n}{N}}_{\frac{1}{N}}d\xi\,f(\xi)
+\frac12\left\{f\left(\frac{1}{N}\right)+f\left(\frac{n}{N}\right)\right\} +\cO(N^{-1})
\ee
converts the sums to integrals. After calculating the integrals, we end up with 
\bea
P^{(H)}_n(x) & = & (\mbox{sgn}(x))^n\,\left(\frac{|x|+\sqrt{x^2-4\frac{n}{N}}}{2}\right)^{n+\frac12}
\frac{1}{(x^2-4\frac{n}{N})^{1/4}} \nn \\
& & \times \exp\left[\frac{N}{4}x^2-\frac{N}{4}|x|\sqrt{x^2-4\frac{n}{N}}-\frac{1}{2}n +\cO(N^{-1})\right]
\label{1-inst_PHn}
\eea
for $|x|>2$. 

By using (\ref{SRh_Hn}) and (\ref{1-inst_PHn}), 
the leading contribution to the correction $\tilde{S}_n$ in (\ref{Sn}):  
\bea
\tilde{S}_n^{(1)} & = & \frac{1}{N}\frac{1}{h^{(H)}_n}\,P^{(H)}_n(-\mu^2)^2\,e^{-\frac{N}{2}\mu^4} 
\label{Stilde_1n}
\eea
can be expressed as 
\bea
\tilde{S}_n^{(1)} & = & \frac{1}{\sqrt{2\pi}}\,\exp\Bigg[\left(n-\frac12\right)\ln N-\ln n! 
+(2n+1)\ln\frac{\mu^2+\sqrt{\mu^4-4\frac{n}{N}}}{2} \nn \\
& & \hspace{18mm}  -\ln\sqrt{\mu^4-4\frac{n}{N}} -\frac{N}{2}\mu^2\sqrt{\mu^4-4\frac{n}{N}} -n\Bigg] 
\times\{1+\cO(N^{-1})\}. 
\label{Stilde_1n2}
\eea
Let us consider the corresponding contribution to (\ref{vevB_S}): 
\be
\left.\vev{\frac{1}{N}\tr\,(\phi^2-\mu^2)}^{(1,0)}\right|_{\rm 1-inst.}=\frac{1}{N}\sum_{n=0}^{N-1}\tilde{S}^{(1)}_n,  
\label{1-inst_vevB}
\ee
where the summand (\ref{Stilde_1n2}) near $n=N-1$ has the mildest damping in the exponent and 
gives a dominant effect, by noting 
\be
\sqrt{\mu^4-4\frac{n}{N}}=2\sqrt{\left(1-\frac{n}{N}\right)+4\omega +4\omega^2}
\ee
with $\omega$ small~\footnote{
For instance, we can see how the summand damps at $n$ away from $n=N-1$ as follows. 
It is easy to find a damping factor $e^{-2N}$ in (\ref{Stilde_1n2}) at $n=\cO(N^0)$.
For $n=aN\gg 1$ with a fractional number $a$ satisfying $1-a\gg \omega$, it turns out that the summand has an exponential damping 
$e^{-\sigma(a)\,N}$. 
Here the function 
$\sigma(x)\equiv 2\sqrt{1-x}-2x\ln(1+\sqrt{1-x}) +x\ln x$ monotonically decreases for $0<x<1$ 
and has the limits: $\lim_{x\to 0}\sigma(x)=2$ and $\lim_{x\to 1}\sigma(x)=0$. 
}. 
Thus, we may consider contributions around the upper limit of the sum (\ref{1-inst_vevB}) and recast it as 
\be
\left.\vev{\frac{1}{N}\tr\,(\phi^2-\mu^2)}^{(1,0)}\right|_{\rm 1-inst.}=\frac{1}{N}\frac{1}{2\pi}\int^{1-\frac{1}{N}}d\xi\,
e^{Nf_0(\xi)+f_1(\xi)}\,\times [1+\cO(N^{-1})]
\label{1-inst_vevB2}
\ee
with 
\bea
f_0(\xi) &\equiv & -\xi\ln \xi +2\xi\ln \frac{\mu^2+\sqrt{\mu^4-4\xi}}{2}-\frac12\mu^2\sqrt{\mu^4-4\xi}, \nn \\
f_1(\xi)& \equiv & -\frac12\ln\xi +\ln\frac{\mu^2+\sqrt{\mu^4-4\xi}}{2}-\ln\sqrt{\mu^4-4\xi}.  
\label{f0f1}
\eea
We take $\xi=(1-\frac{1}{N})\xi'$ and change the integration variable to $s=\sqrt{1-\frac{4\xi'}{\mu^4}}$ 
to obtain
\be
\left.\vev{\frac{1}{N}\tr\,(\phi^2-\mu^2)}^{(1,0)}\right|_{\rm 1-inst.}=\frac{1}{N}\frac{-\mu^2}{4\pi}\,
F\left(\sqrt{1-\frac{4}{\mu^4}}\right)\times [1+\cO(N^{-1})], 
\ee
where 
\bea
F(\varepsilon) & \equiv & \int^{\varepsilon}ds\,\exp\left[N\mu^4\left\{\frac{1-s^2}{4}\ln\left(1+\frac{2s}{1-s}\right)-\frac12s\right\}\right] \nn \\
& &  \hspace{8mm} \times \exp\left[-\frac{\mu^4(1-s^2)-2}{4}\ln \left(1+\frac{2s}{1-s}\right) \right]. 
\eea
Because $\sqrt{1-\frac{4}{\mu^4}}=2\sqrt{\omega}\left[1-\frac32\omega+\cO(\omega^2)\right]$ is a small 
quantity, let us 
consider $F(\varepsilon)$ for $\varepsilon$ small. 
When $N\varepsilon^3$ is kept finite but large as $N\sim \infty$ and $\varepsilon\sim 0$, 
\be
F(\varepsilon) = \int^{\varepsilon}ds\,e^{-N\mu^4\,\frac{s^3}{3}}\,(1+\cO(s)) 
= \frac{-1}{N\mu^4\varepsilon^2}\,e^{-N\mu^4\,\frac{\varepsilon^3}{3}}\,
\left[1+\cO\left(\frac{1}{N\varepsilon^{3}}\right)\right] . 
\label{F_epsilon}
\ee
The expression on the r.h.s. is confirmed by taking a derivative with respect to $\varepsilon$. 
This gives the final result 
\be
\left.\vev{\frac{1}{N}\tr\,(\phi^2-\mu^2)}^{(1,0)}\right|_{\rm 1-inst.} =  
N^{-4/3}\,\hat{\Omega}^{(1)}_0(t)\,\left[1+\cO(t^{-3/2})\right]
\label{1-inst_vevBf} 
\ee
with 
\be
\hat{\Omega}^{(1)}_0(t) \equiv \frac{1}{32\pi t}\,e^{-\frac{32}{3}\,t^{3/2}} 
\label{Omega0(1)hat}
\ee
for $t$ fixed to be finite but large. 
From (\ref{vevB_Z_alpha}) and (\ref{1-inst_vevBf}), we can conclude that the nonperturbative effect 
dynamically breaks the supersymmetry 
(under wave function renormalization absorbing the factor $N^{-4/3}$)~\footnote{
The wave function renormalization can be understood from the finite expression for the free 
energy (\ref{1-inst_F10}). The renormalized one-point function is given by the $t$-derivative of the 
free energy multiplied by the factor $(-\frac14)$. Note (\ref{vevB_Z10}) and 
the relation $\mu^2=2+4\omega$.}. 

The contribution to the free energy $F_{(1,0)}$ is obtained by integrating 
(\ref{vevB_Z10}) as 
\bea
\left.F_{(1,0)}\right|_{\rm 1-inst.} & = & 4\int^{\infty}_t dt'\,\hat{\Omega}^{(1)}_0(t')\,
\left[1+\cO(t'^{-3/2})\right] \nn \\
& = & \frac{1}{128\pi\,t^{3/2}}\,e^{-\frac{32}{3}\,t^{3/2}}\,
\left[1+\cO(t^{-3/2})\right].
\label{1-inst_F10}
\eea
The integration constant is determined from the fact that there is no perturbative contribution to $F_{(1,0)}$,
as seen from (\ref{Z10_localization}) or (\ref{Z0inst}). 
Notice that
(\ref{1-inst_F10}) coincides with the result (\ref{Fenergy_f}) not only in the exponential factor 
$e^{-\frac{32}{3}\,t^{3/2}}$ 
but also in the prefactor $\frac{1}{128\pi\,t^{3/2}}$. 
Moreover, the agreement of the exponential factors is already seen before taking the double scaling 
limit. Namely, $(V_{\rm eff}^{(0)}(0)-1)$ obtained from (\ref{Veff0at0}) is 
exactly equal to $-f_0(1)$ from (\ref{f0f1}). 
It gives additional grounds for regarding the results (\ref{1-inst_vevBf}) and (\ref{1-inst_F10}) 
as one-instanton contributions to $\vev{\frac{1}{N}\tr\,(\phi^2-\mu^2)}^{(1,0)}$ and $F_{(1,0)}$, respectively. 
Thus (\ref{1-inst_vevBf}) shows that the instanton induces the supersymmetry breaking.

\subsection{Full one-instanton contribution} 
\label{sec:1-inst_full}
Here we compute the one-instanton effect to all orders, 
namely full contributions to the factors $[1+\cO(t^{-3/2})]$ in (\ref{1-inst_vevBf}) and (\ref{1-inst_F10}). 

Substituting (\ref{PH}), (\ref{SRh_Hn}) and (\ref{Stilde_1n}) in (\ref{1-inst_vevB}), we have 
\be
\left.\vev{\frac{1}{N}\tr\,(\phi^2-\mu^2)}^{(1,0)}\right|_{\rm 1-inst.}=
\frac{e^{-z^2}}{\sqrt{2\pi}\,N^{3/2}}\,\frac{1}{2^N \,(N-1)!}\,
\left[H_N(z)^2-H_{N-1}(z)\,H_{N+1}(z)\right], 
\label{1-install_vevB}
\ee
where 
\be
z\equiv \sqrt{\frac{N}{2}}\,\mu^2=\sqrt{2N}\,(1+2\omega),  
\ee
and the relation 
\be
\sum_{k=0}^{n-1}\frac{1}{2^k k!}\,H_k(x)^2=\frac{1}{2^n\,(n-1)!}\left[H_n(x)^2-H_{n-1}(x)\,H_{n+1}(x)\right]
\ee
was used. The latter can be proved by an inductive argument. 
Upon taking the double scaling limit in (\ref{1-install_vevB}), 
the following asymptotic formula plays a relevant role: 
\be
e^{-x^2/2}H_n(x) = \pi^{\frac14} 2^{\frac{n}{2}+\frac14} n^{-\frac{1}{12}}\sqrt{n!}\left[{\rm Ai}(s) 
+ n^{-2/3}g_0(s) +n^{-1}g_1(s)+\cdots\right] 
\label{hermite_asymp}
\ee 
which is valid for large $n$ with 
\be
x=\sqrt{2n+1}+\frac{s}{\sqrt{2}\,n^{1/6}}. 
\label{x_s}
\ee
The Airy function is defined by 
\be
{\rm Ai}(s)\equiv \frac{1}{2\pi}\int^\infty_{-\infty} dz\,e^{-isz-\frac{i}{3}z^3}, 
\label{airy_def}
\ee
and $g_0(s), g_1(s), \cdots$ are functions depending only on $s$. 
The appearance of the Airy function in (\ref{hermite_asymp}) seems reasonable from the WKB analysis of the 
harmonic oscillator potential around its turning points. 
See appendix~\ref{app:formula} for a derivation of (\ref{hermite_asymp}). 
In applying (\ref{hermite_asymp}) to (\ref{1-install_vevB}), notice that 
\bea
s & = & 4t-\frac12N^{-1/3}+\frac{1}{16}N^{-4/3}+\cO(N^{-7/3})\qquad \mbox{for} \quad H_N, \nn \\
s & = & 4t+\frac12N^{-1/3}-\frac23 N^{-1}t-\frac{1}{48}N^{-4/3}+\cO(N^{-2})\qquad \mbox{for} \quad H_{N-1}, \nn \\
s & = & 4t-\frac32N^{-1/3}+\frac23 N^{-1}t+\frac{5}{16}N^{-4/3}+\cO(N^{-2})\qquad \mbox{for} \quad H_{N+1}.
\eea
We find that the full one-instanton contribution is given in terms of the Airy function and its derivative, by 
\be
\left.\vev{\frac{1}{N}\tr\,(\phi^2-\mu^2)}^{(1,0)}\right|_{\rm 1-inst.} =  
N^{-4/3}\left[\Omega_0^{(1)}(t) +N^{-2/3}\,\Omega^{(1)}_{2/3}(t)+\cO(N^{-1})\right], \\
\label{1-install_vevBf}
\ee
with
\be
\Omega_0^{(1)}(t) \equiv {\rm Ai}'(4t)^2-4t{\rm Ai}(4t)^2
\label{Omega0(1)}
\ee
and corrections at order $N^{-2/3}$ nonvanishing. The subleading term 
\be
\Omega^{(1)}_{2/3}(t) 
= -\frac{12}{5}t^2{\rm Ai}(4t)^2+\frac{3}{20}\,{\rm Ai}(4t){\rm Ai}'(4t)+\frac25t\,{\rm Ai}'(4t)^2
\label{Omega1_2/3}
\ee
is obtained by using (\ref{g0}). 
From the asymptotic expansion of the Airy function: 
\be
{\rm Ai}(s)=\frac{1}{2\pi}\frac{1}{s^{1/4}}\,e^{-\frac23 s^{3/2}}\sum_{n=0}^\infty 
\frac{\Gamma(\frac12+3n)}{(2n)!}\left(-\frac{1}{9s^{3/2}}\right)^n
\ee
for large $s$, we see that all-order corrections to (\ref{Omega0(1)hat}) take the form of 
\be
\Omega_0^{(1)}(t) =  
\frac{1}{32\pi t}\,e^{-\frac{32}{3}t^{3/2}}\,
\left[1+\sum_{n=1}^\infty a^{(1)}_n\,\frac{1}{t^{\frac32n}}\right] 
\label{1-install_vevB_series}
\ee
with $a^{(1)}_1=-\frac{17}{192}, \,a^{(1)}_2=\frac{1225}{73728}, \,a^{(1)}_3=-\frac{199115}{42467328}, \cdots$. 
The power series with respect to $t^{-3/2}$ 
can be regarded as perturbative contributions to all orders  
around the one-instanton configuration~\footnote{
We can systematically improve the r.h.s. of (\ref{F_epsilon}) and obtain a series 
\be
\left.\vev{\frac{1}{N}\tr\,(\phi^2-\mu^2)}^{(1,0)}\right|_{\rm 1-inst.}= N^{-4/3}\,
\frac{1}{32\pi t}\,e^{-\frac{32}{3}\,t^{3/2}}\,\left[\sum_{n=0}^\infty \frac{\Gamma(n+\frac23)}{\Gamma(\frac23)}\left(-\frac{3}{32\,t^{3/2}}\right)^n\right]. 
\ee
However, it does not coincide with (\ref{1-install_vevB_series}). Presumably, 
higher order contributions in $1/N$ to (\ref{PHn}) or (\ref{1-inst_vevB2}) 
which were omitted in section~\ref{sec:1-inst_leading} could yield nonvanishing 
contributions in the double scaling limit, and would account for the difference. 
Recalling (\ref{dsl}), if a term of order $N^{-1}$ in the last factor $[1+\cO(N^{-1})]$ in (\ref{1-inst_vevB2}) appears together with 
$\omega^{-3/2}$, it gives rise to $t^{-3/2}$ which contributes in the double scaling limit. 
In general, in order to reproduce $a^{(1)}_n$ in (\ref{1-install_vevB_series}), 
contributions of order $N^{-n}$ would have to be taken into account in the factor $[1+\cO(N^{-1})]$ 
in (\ref{1-inst_vevB2}).
}. 
Similar to (\ref{1-inst_F10}), the full one-instanton contribution to the free energy $F_{(1,0)}$ gives
\bea
\left.F_{(1,0)}\right|_{\rm 1-inst.} & = & 4\int^{\infty}_t dt' \,\Omega_0^{(1)}(t') \nn \\
& = & \frac13\left[32t^2\,{\rm Ai}(4t)^2-{\rm Ai}(4t)\,{\rm Ai}'(4t)-8t\,{\rm Ai}'(4t)^2\right]
\label{1-inst_F10_airy} \\
& = & \frac{1}{128\pi\,t^{3/2}}\,e^{-\frac{32}{3}\,t^{3/2}}\,
\left[1+\sum_{n=1}^\infty b^{(1)}_n\,\frac{1}{t^{\frac32n}}\right]
\label{1-inst_F10_series}
\eea
with $b^{(1)}_1=-\frac{35}{192}, \,b^{(1)}_2=\frac{3745}{73728}, \,b^{(1)}_3=-\frac{805805}{42467328}, \cdots$.

Interestingly, (\ref{Omega0(1)}) and (\ref{1-inst_F10_airy}) are closed form 
expressions and include fluctuations to all orders around the one-instanton configuration. 
The justification for this claim will become more evident in the next section, 
where we observe that all additional contributions to the one-point function and free energy involve 
only higher powers of $e^{-\frac{32}{3}\,t^{3/2}}$, 
and are thus attributed to $k$-instantons with $k>1$. 
It is an intriguing aspect of our supersymmetric matrix model, 
because in matrix models for bosonic strings 
such an expression has not been obtained even for the simplest case of $c=0$. 
It would be interesting to investigate the large order behavior of $a^{(1)}_n$ in 
(\ref{1-install_vevB_series}) or $b^{(1)}_n$ in (\ref{1-inst_F10_series}) and to compare 
the result with the large order growth $(2n)!$ in a string perturbation series~\cite{Shenker:1990uf}. 
Knowledge of this behavior could provide some insight into the stability of the one-instanton background.

\section{Leading order two-instanton contribution}
\label{sec:2-inst}
\setcounter{equation}{0}
In this section, we calculate the leading order two-instanton contribution to the one-point function and free 
energy. 
For the effect on the one-point function (\ref{vevB_S}), we need to know $\tilde{S}^{(2)}_n$. 

\subsection{Calculation of $\tilde{S}^{(2)}_n$}
{}From (\ref{Sn}), (\ref{PSRh_H_tilde}) and (\ref{Stilde_1n}), one finds
\be
\tilde{S}_n ^{(2)} = \tilde{S}_n^{(1)}\left[-\frac{\tilde{h}_n^{(1)}}{h_n^{(H)}} 
+2\tilde{L}_n^{(1)}(-\mu^2)\right].
\label{tilS^2}
\ee
In order to compute $\tilde{L}_n^{(1)}(-\mu^2)$, 
we start from the first order expression for (\ref{recursion_1st_k}) obtained in an instanton number expansion: 
\be
-k_{m+1}^{(H)}(x) \,\tilde{L}^{(1)}_{m+1}(x) +x\tilde{L}^{(1)}_m(x) 
-\frac{\xi}{k_m^{(H)}(x)}\,\tilde{L}^{(1)}_{m-1}(x) 
 =\left(1+\frac{\xi}{k_m^{(H)}(x)}\frac{1}{k^{(H)}_m(-\mu^2)}\right)\tilde{S}^{(1)}_m \ ,
 \label{recursion_1st_k_L}
\ee
with $\xi=\frac{m}{N}$. 
Here, 
\be
\tilde{R}_m^{(1)} = \frac{1}{N}\frac{1}{h_{m-1}^{(H)}}P_m^{(H)}(-\mu^2)P_{m-1}^{(H)}(-\mu^2)\,e^{-\frac{N}{2}\mu^4} =\frac{\xi}{k_m^{(H)}(-\mu^2)}\,\tilde{S}_m^{(1)} 
\label{tilR_m}
\ee
was used. Since (\ref{tilR_m}) can also be expressed as $k^{(H)}_m(-\mu^2)\,\tilde{S}_{m-1}^{(1)}$, 
the relation 
\be
\tilde{S}^{(1)}_{m-1}=\frac{\xi}{k_m^{(H)}(-\mu^2)^2}\,\tilde{S}_m^{(1)} 
\label{tilS_m_m-1}
\ee
is obtained. 
For the leading order term in the two-instanton contribution, we may plug (\ref{kH}), (\ref{k0k1}) and 
\be
\tilde{S}_m^{(1)} = \frac{1}{2\pi}\frac{1}{N}\,e^{Nf_0(\xi)+f_1(\xi)}\,\times [1+\cO(N^{-1})]
\label{tilS_m^1}
\ee
with (\ref{f0f1}) into the recursion relation (\ref{recursion_1st_k_L}). 
We find a solution for (\ref{recursion_1st_k_L}) by assuming the form of $\tilde{L}^{(1)}_m(x)$ as 
\be
\tilde{L}_m^{(1)}(x) = \left[L(x,\xi) + \cO(N^{-1})\right] \tilde{S}_m^{(1)}. 
\label{assume_Lm}
\ee
Namely, $L(x, \xi)$ depends on $N$ only through $\xi$, and $\tilde{L}^{(1)}_m(x)$ and $\tilde{S}^{(1)}_m$ 
are of the same order in $1/N$.  

By using (\ref{tilS_m_m-1}), the recursion relation at leading order in $1/N$ becomes 
\bea
 & & \left[-k^{(0)}(x,\xi)\frac{1}{\xi}k^{(0)}(-\mu^2,\xi)^2+x-\frac{1}{k^{(0)}(x,\xi)}\xi^2\frac{1}{k^{(0)}(-\mu^2,\xi)^2}\right]L(x,\xi) 
\nn \\
& & = 1+\frac{1}{k^{(0)}(x,\xi)}\frac{\xi}{k^{(0)}(-\mu^2,\xi)}, 
\eea
from which we have 
\be
L(x,\xi) = \frac{k^{(0)}(x,\xi)-k^{(0)}(-\mu^2,\xi)-\mu^2}{xk^{(0)}(x,\xi)\left(2+\frac{\mu^2}{\xi}k^{(0)}(-\mu^2,\xi)\right) -2\mu^2 k^{(0)}(-\mu^2,\xi)-\mu^4}. 
\label{sol_L}
\ee
This reduces to a simple formula at $x=-\mu^2$: 
\be
L(-\mu^2,\xi)= \frac{\mu^2-\sqrt{\mu^4-4\xi}}{2(\mu^4-4\xi)}, 
\label{sol_L_2}
\ee
and thus we obtain
\be
\tilde{L}_m^{(1)}(-\mu^2) = \left[\frac{\mu^2-\sqrt{\mu^4-4\frac{m}{N}}}{2(\mu^4-4\frac{m}{N})}+\cO(N^{-1})\right]
\,\tilde{S}_m^{(1)} \ .
\label{sol_Lm}
\ee

Next, let us obtain $\tilde{h}_n^{(1)}$ in (\ref{tilS^2}). 
From (\ref{PSRh_H_tilde}) one may rewrite $h_n =\left(\prod_{m=1}^n R_m\right)h_0$ as 
\bea
h_n  & = & \left(\prod_{m=1}^n R_m^{(H)}\right)\prod_{m=1}^n\left(1+\frac{\tilde{R}_m}{R_m^{(H)}}\right)\cdot 
h_0^{(H)}\left(1+\frac{\tilde{h}_0}{h_0^{(H)}}\right) \nn \\
& = & h_n^{(H)} \left[1+\frac{\tilde{h}_0}{h_0^{(H)}}+\sum_{m=1}^n\frac{\tilde{R}^{(1)}_m}{R_m^{(H)}} + \mbox{(higher orders)}\right],  
\eea
from which we read off 
\be
\frac{\tilde{h}_n^{(1)}}{h_n^{(H)}}=\frac{\tilde{h}_0}{h_0^{(H)}}+\sum_{m=1}^n\frac{N}{m}\,\tilde{R}^{(1)}_m. 
\label{hn_ratio}
\ee
Note that (\ref{h0}) and (\ref{SRh_Hn}) allow us to express the first term in terms of the error function: 
\bea
\frac{\tilde{h}_0}{h_0^{(H)}} & = & -\frac{1}{2}\,\mbox{erfc}\left(\sqrt{\frac{N}{2}}\mu^2\right)
\nn \\
 & = & -\frac{1}{\sqrt{2\pi N}}\frac{1}{\mu^2}\,e^{-\frac{N}{2}\mu^4}\,[1+\cO(N^{-1})]. 
 \label{erfc}
\eea
Substituting (\ref{sol_Lm}) and (\ref{hn_ratio}) into (\ref{tilS^2}), we have 
\bea
\tilde{S}_n^{(2)} & = & \tilde{S}_n^{(1)} \frac{1}{2}\,\mbox{erfc}\left(\sqrt{\frac{N}{2}}\mu^2\right)
\nn \\
 & & +\tilde{S}_n^{(1)} \sum_{m=1}^n\frac{\mu^2-\sqrt{\mu^4-4\frac{m}{N}}}{2\frac{m}{N}}\,\tilde{S}_m^{(1)} +\frac{\mu^2-\sqrt{\mu^4-4\frac{n}{N}}}{\mu^4-4\frac{n}{N}}\left(\tilde{S}^{(1)}_n\right)^2
 \label{tilS^2_1}
\eea
after use of (\ref{tilR_m}) and (\ref{erfc}). 

\subsection{Calculation of nonperturbative effects}

The sum $\frac{1}{N}\sum_{n=0}^{N-1}\tilde{S}^{(2)}_n$ gives the two-instanton contribution to  $\vev{\frac{1}{N}\tr\,(\phi^2-\mu^2)}^{(1,0)}$. 
The first term in the summand (\ref{tilS^2_1}) is negligible at large $N$ due to the exponential damping in the error function. 
We therefore focus on the region $m\sim n\sim N-1$ in the sum of the second and third terms, which give 
relevant contributions. 

\paragraph{The second term}
Let us first consider the sum 
\be
\sum_{m=1}^n\frac{\mu^2 -\sqrt{\mu^4-4\frac{m}{N}}}{2\frac{m}{N}}\,\tilde{S}_m^{(1)}
\label{2ndterm}
\ee
in the second term, focusing on the region $m = N-\cO(N^0)$, $n=N-\cO(N^0)$.  
Use of (\ref{tilS_m^1}) gives 
\be
(\ref{2ndterm}) 
= \frac{1}{2\pi}\int^\xi d\eta \,\frac{\mu^2-\sqrt{\mu^4-4\eta}}{2\eta}\,e^{Nf_0(\eta)+f_1(\eta)} \, \times
\left[1+\cO(N^{-1})\right]. 
\ee 
Here, we take $\eta=\frac{m}{N}$ and $\xi=\frac{n}{N}$. 
We change the integration variable to $\eta = \xi\eta' = \eta' -(1-\xi)\eta'$, and expand functions of 
$\eta'-(1-\xi)\eta'$ around $\eta'$. By noting $1-\xi=\cO(N^{-1})$, we find
\be
(\ref{2ndterm}) =  \frac{1}{2\pi}\int^1d\eta'\,\frac{\mu^2-\sqrt{\mu^4-4\eta'}}{2\eta'\sqrt{\mu^4-4\eta'}} 
\, e^{Nf_0(\eta')+\tilde{f}_1(\eta')}
\times [1+\cO(N^{-1})], 
\ee
where 
\be
\tilde{f}_1(\eta')\equiv \left(N(1-\xi)\eta'-\frac12\right)\left\{\ln \eta' 
-2\ln \frac{\mu^2+\sqrt{\mu^4-4\eta'}}{2}\right\}. 
\ee
Next, the variable change $s=\sqrt{1-\frac{4\eta'}{\mu^4}}$ leads to 
\be
 (\ref{2ndterm}) =  \frac{-1}{2\pi}F_1\left(\sqrt{1-\frac{4}{\mu^4}}\right)\, \times[1+\cO(N^{-1})]
\ee
with 
\bea
F_1(\varepsilon) & \equiv &\int^\varepsilon\frac{ds}{1+s}\, 
 \exp\left[N\mu^4\left\{\frac{1-s^2}{4}\ln\left(1+\frac{2s}{1-s}\right)-\frac12 s\right\}\right] \nn \\
 & &\times 
\exp\left[-\left\{N(1-\xi)\frac{\mu^4}{4}(1-s^2)-\frac12\right\}\ln\left(1+\frac{2s}{1-s}\right)\right].
 \eea
Similar to (\ref{F_epsilon}), 
\be
F_1(\varepsilon) = \int^{\varepsilon} ds\,e^{-N\mu^4\frac{s^3}{3}}\, (1+\cO(s)) 
= \frac{-1}{N\mu^4\varepsilon^2}\,e^{-N\mu^4\frac{\varepsilon^3}{3}}\,
\left[1+\cO\left(\frac{1}{N\varepsilon^{3}}\right)\right]
\label{F1_epsilon}
\ee 
is obtained for $N\varepsilon^3$ finite but large. Using this result gives
\be
(\ref{2ndterm}) 
= N^{-1/3}\,\frac{1}{32\pi t}\,e^{-\frac{32}{3}\,t^{3/2}}\,\left[1+\cO(t^{-3/2})\right]. 
\label{sum_2ndterm2}
\ee
Since the $\xi$-dependent part does not contribute to (\ref{sum_2ndterm2}), 
the summation of the second term in (\ref{tilS^2_1}) with respect to $n$ reduces to $\frac{1}{N}\sum_{n=0}^{N-1}\tilde{S}_n^{(1)}$, which
is nothing but (\ref{1-inst_vevBf}). 
Thus we arrive at 
\be
\frac{1}{N}\sum_{n=0}^{N-1}\tilde{S}_n^{(1)} \sum_{m=1}^n\frac{\mu^2 -\sqrt{\mu^4-4\frac{m}{N}}}{2\frac{m}{N}}\,\tilde{S}_m^{(1)} 
= N^{-5/3}\left(\frac{1}{32\pi t}\right)^2\,e^{-\frac{64}{3}\,t^{3/2}}\,\left[1+\cO(t^{-3/2})\right]. 
\label{sum_2ndterm_f}  
\ee

\paragraph{The third term}
The sum of the third term 
\be
\frac{1}{N}\sum_{n=0}^{N-1}\frac{\mu^2-\sqrt{\mu^4-4\frac{n}{N}}}{\mu^4-4\frac{n}{N}}\left(\tilde{S}^{(1)}_n\right)^2
\label{sum_3rdterm}
\ee
can be evaluated in a similar manner. 
We convert the summation to an integral by taking $\xi=\frac{n}{N}$ as 
$\frac{1}{N}\sum_{n=0}^{N-1} \to \int^{1-\frac{1}{N}} d\xi$, and 
change the variable to $\xi=(1-\frac{1}{N})\xi'$.
This is then followed by a further variable change $s=\sqrt{1-\frac{4\xi'}{\mu^4}}$. 
The result is 
\be
(\ref{sum_3rdterm}) = \frac{-1}{8\pi^2}\frac{1}{N^2\mu^2}F_2\left(\sqrt{1-\frac{4}{\mu^4}}\right)\, \times[1+\cO(N^{-1})]
\ee
with 
\bea
F_2(\varepsilon) & \equiv &\int^\varepsilon ds\,\frac{1-s}{s^3}\, 
 \exp\left[N\mu^4\left\{\frac{1-s^2}{2}\ln\left(1+\frac{2s}{1-s}\right)- s\right\}\right] \nn \\
 & &\hspace{19mm}\times 
\exp\left[-\left\{\frac{\mu^4}{2}(1-s^2)-1\right\}\ln\left(1+\frac{2s}{1-s}\right)\right].
\eea
When $N\varepsilon^3$ is finite but large, $F_2(\varepsilon)$ can be evaluated as 
\be
F_2(\varepsilon) = \frac{-1}{2N\mu^4\varepsilon^5}\,e^{-2N\mu^4\frac{\varepsilon^3}{3}}\,
\left[1+\cO\left(\frac{1}{N\varepsilon^{3}}\right)\right], 
\ee
leading to the result 
\be
\frac{1}{N}\sum_{n=0}^{N-1}\frac{\mu^2-\sqrt{\mu^4-4\frac{n}{N}}}{\mu^4-4\frac{n}{N}}\left(\tilde{S}^{(1)}_n\right)^2 
 =N^{-4/3}\,\frac{1}{(64\pi)^2 \,t^{5/2}}\, e^{-\frac{64}{3}\,t^{3/2}}\,
\left[1+\cO(t^{-3/2})\right]. 
\label{sum_3rdterm_f}
\ee
 
\paragraph{Final result}
Comparing the powers of $N$ between (\ref{sum_2ndterm_f}) and (\ref{sum_3rdterm_f}), 
we find that the latter is dominant. 
Thus, the two-instanton contribution is found to be
\be
\left.\vev{\frac{1}{N}\tr\,(\phi^2-\mu^2)}^{(1,0)}\right|_{\rm 2-inst.} =
\frac{1}{N}\sum_{n=0}^{N-1}\tilde{S}_n^{(2)} = N^{-4/3}\,\hat{\Omega}^{(2)}_0(t)\,
\left[1+\cO(t^{-3/2})\right] 
\label{2-inst_vevBf} 
\ee
with
\be
\hat{\Omega}^{(2)}_0(t) \equiv \frac{1}{(64\pi)^2 \,t^{5/2}}\, e^{-\frac{64}{3}\,t^{3/2}},
\label{Omega0(2)hat}
\ee
for $t$ finite but large. 
Both effects from one instanton (\ref{1-inst_vevBf}) and from two instantons (\ref{2-inst_vevBf}) are of the same order in $N$ 
and equally contribute in the double scaling limit to the quantity 
\be
N^{4/3}\,\vev{\frac{1}{N}\tr\,(\phi^2-\mu^2)}^{(1,0)}=N^{4/3}\,\frac{1}{N}\sum_{n=0}^{N-1}
\left\{\tilde{S}_n^{(1)}+\tilde{S}_n^{(2)}+\cdots\right\}. 
\ee
The weight of the exponential in (\ref{Omega0(2)hat}) is twice that of (\ref{Omega0(1)hat}), 
as it should be from the interpretation of a two-instanton contribution.  
In general, $\Omega^{(k)}_0(t)$ denotes the double scaling limit of 
$N^{4/3}\,\frac{1}{N}\sum_{n=0}^{N-1}\tilde{S}^{(k)}_n$ 
and $\hat{\Omega}_0^{(k)}(t)$ its leading large-$t$ behavior; these are both expected to 
give $k$-instanton contributions scaling as $e^{-\frac{32k}{3}\,t^{3/2}}$. 
Notice that we do not use the dilute gas approximation for instantons in this calculation. Namely, 
interactions among instantons are taken into account.  
 
Correspondingly, the free energy is expressed as 
\be
F_{(1,0)}=\left. F_{(1,0)}\right|_{\rm 1-inst.} + \left.F_{(1,0)}\right|_{\rm 2-inst.} +\cdots, 
\ee
where the first term is the one-instanton contribution given by (\ref{1-inst_F10_airy}) or (\ref{1-inst_F10_series}), and the second term is 
\be
\left.F_{(1,0)}\right|_{\rm 2-inst.}=\frac12\frac{1}{(128\pi)^2\,t^3} \,e^{-\frac{64}{3}\,t^{3/2}}
\,\left[1+\cO(t^{-3/2})\right] 
\label{2-inst_F10}
\ee
due to two instantons. 
Since the dilute gas approximation does not give rise to multi-instanton contributions to the free energy~\footnote{
{}From (\ref{Fenergy}), disconnected multi-instanton amplitudes do not contribute to the free energy.}, 
the result (\ref{2-inst_F10}) is considered to be attributed solely to interactions between the instantons.

Since the contribution from each instanton sector is expected to be equally important in the double scaling limit, 
it would be nontrivial whether 
the full result including all instanton contributions gives a well-defined quantity or not, in particular for 
small $t$. 
In the next section, we will see numerical evidence suggesting that it is indeed well-defined.

\section{Numerical result for full nonperturbative effects}
\label{sec:full}
\setcounter{equation}{0}
In this section, we numerically compute the expectation value 
$\vev{\frac{1}{N}\tr\,(\phi^2-\mu^2)}^{(1,0)}$ and the free energy $F_{(1,0)}$ including full nonperturbative 
effects. 
Starting with $P_0(-\mu^2)=1$ and the expressions for $h_0$ and $S_0$ in (\ref{h0}) and (\ref{S0}) for a given value of $N$, 
we can carry out the following iterative procedure, beginning from $n=1$:
\begin{enumerate}
\item
Evaluate $P_n(-\mu^2)$ from (\ref{recursion_P}).
\item
Evaluate $R_n$ from (\ref{Rn}).
\item
Evaluate $h_n$ from (\ref{hn_Rn}). 
\item
Evaluate $S_n$ from (\ref{Sn}). 
\item
Go back to 1. with $n$ incremented by one. 
\end{enumerate}
This procedure is repeated $N-1$ times to evaluate the values for $S_n$ ($n=1,\cdots,N-1$).
The resulting $S_n$ are then combined with (\ref{vevB_S}) to determine the exact one-point function 
for a given $N$ and $t$.

%
\begin{figure}
\centering
\includegraphics[height=7cm, width=12cm, clip]{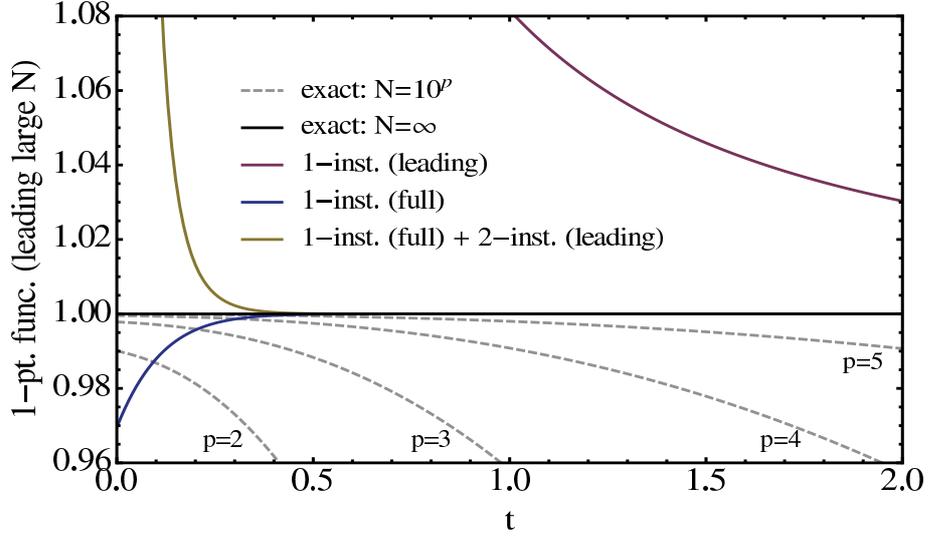}
\caption{$\Omega(N, t)$ defined by (\ref{Omega}) as a function of $t$. 
Everything is normalized by the $N=\infty$ result $\Omega_0(t)\equiv\Omega(\infty, t)$ (exact: $N=\infty$), 
and thus the black solid line representing it is flat. 
The gray dashed lines (exact: $N=10^p$) show the results $\Omega(N, t)$ for $N=10^p$ ($p=2,3,4,5$). 
The red line (1-inst. (leading)) and the blue line (1-inst. (full)) show the behavior
of the leading one-instanton contribution $\hat{\Omega}^{(1)}_0(t)$ in (\ref{Omega0(1)hat}) and 
the full one-instanton contribution $\Omega_0^{(1)}(t)$ in (\ref{Omega0(1)}), respectively. 
Finally, the yellow line (1-inst. (full) + 2-inst. (leading)) represents 
the sum of the full one-instanton result $\Omega_0^{(1)}(t)$ 
and the leading two-instanton result $\hat{\Omega}^{(2)}_0(t)$ in (\ref{Omega0(2)hat}). 
} 
\label{fig:1ptfn}
\end{figure}
%
We evaluate the one-point function 
\be
\Omega(N, t)\equiv N^{4/3}\vev{\frac{1}{N}\tr\,(\phi^2-\mu^2)}^{(1,0)}
\label{Omega}
\ee 
from $N=1$ to $N=1,000,000$, 
and then extrapolate the results to 
$N=\infty$ to obtain $\Omega_0(t)\equiv \Omega(\infty, t)$. 
The systematic error in the extrapolation is expected to be too small (around $10^{-7}\%$) 
to resolve in the presented figures. 
Fig.~\ref{fig:1ptfn} summarizes our result for the one-point function. 
Everything in the plot is normalized by the $N=\infty$ result $\Omega_0(t)$, 
and we can see how finite $N$ results depicted by the gray dashed lines converge 
to the $N=\infty$ one. 
The result suggests that $t$ in (\ref{dsl}) is the appropriate scaling variable in the double scaling limit. 
If this were not the case, $\Omega_0(t)$ would be driven to zero or infinity, and consequently all the gray dashed lines 
would lie at infinity or zero. Then, we could not obtain a sensible result such as in Fig.~\ref{fig:1ptfn}. 
The analytical results obtained in the previous sections are also plotted with this normalization. 
The full one-instanton 
result (\ref{Omega0(1)}) significantly improves the approximation compared with the leading result 
(\ref{Omega0(1)hat}). Although it is not clear in this figure 
whether or not the leading two-instanton contribution (\ref{Omega0(2)hat}) 
makes the situation better, magnifying the neighborhood of 1.00 along the vertical axis as in 
Fig.~\ref{fig:1ptfn_zoom} shows that it is indeed the case for $t\gtrsim 0.65$. 
%
\begin{figure}
\centering
\includegraphics[height=7cm, width=12cm, clip]{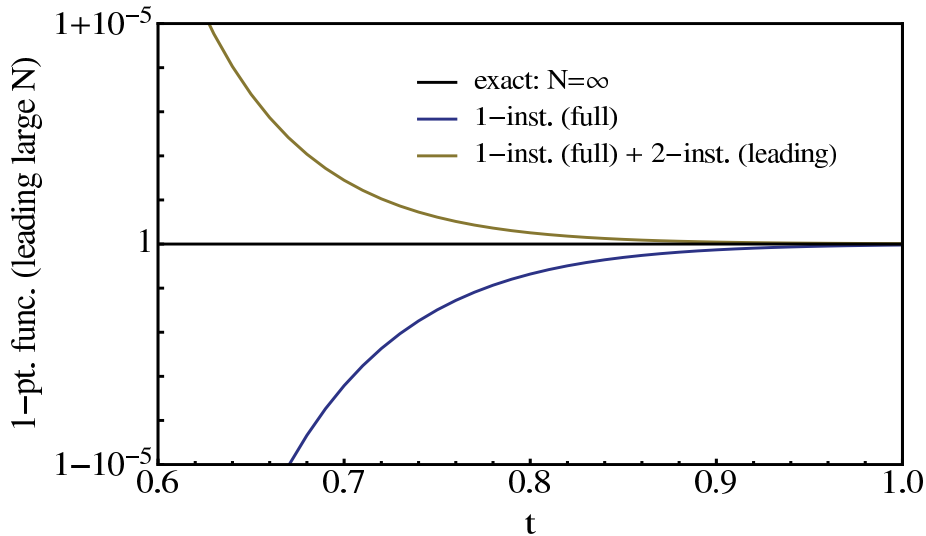}
\caption{A magnified view of Fig.~\ref{fig:1ptfn} around 1.00 in the vertical axis. Finite $N$ results lie outside the plot range. } 
\label{fig:1ptfn_zoom}
\end{figure}
%

We also find that the subleading correction with respect to $1/N$ (\ref{Omega1_2/3}) 
has good agreement with the corresponding numerical result, which is obtained from the subleading extrapolation parameters in $1/N$. 
For readers who have an interest, we present the result in Fig.~\ref{fig:1ptfn_subleadingN} of appendix~\ref{app:1ptfn_subleadingN}.

%
\begin{figure}
\centering
\includegraphics[height=7cm, width=12cm, clip]{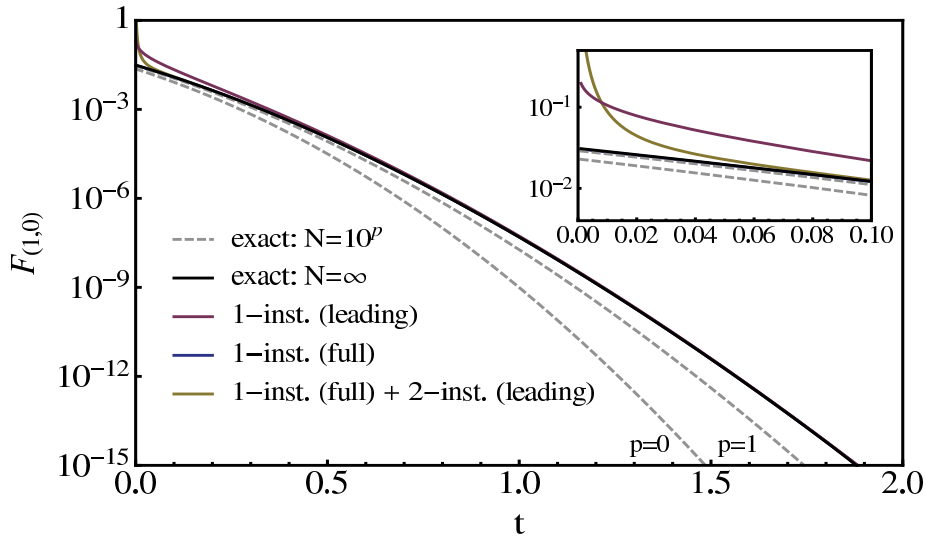}
\caption{Full nonperturbative contribution to the free energy $F_{(1,0)}$ as a function of $t$. 
The black solid line (exact: $N=\infty$) represents the result of (\ref{FreeEnergy_1ptfn}).  
For comparison, finite $N$ results ($4\int^{\infty}_{t} dt'\,\Omega(N,t')$ for various $N$) 
are shown by the gray dashed lines (exact: $N=10^p$).  
Also, the leading and full one-instanton contributions to $F_{(1,0)}$ 
(\ref{1-inst_F10}) and (\ref{1-inst_F10_airy}) are depicted by the red and blue lines, 
respectively. The yellow line represents the sum of the full one-instanton result (\ref{1-inst_F10_airy}) 
and the leading two-instanton result in (\ref{2-inst_F10}). } 
\label{fig:FreeEnergy}
\end{figure}
%
Finally we present in Fig.~\ref{fig:FreeEnergy} the full nonperturbative contribution 
to the free energy $F_{(1,0)}=-\ln Z_{(1,0)}$ obtained 
by numerically integrating the $N=\infty$ result for the one-point function:  
\be
F_{(1,0)}=4\int^{\infty}_{t} dt'\,\Omega_0(t'). 
\label{FreeEnergy_1ptfn}
\ee
Note that the free energy is a finite function of $t$ even at the origin, 
which corresponds to the strongly coupled limit 
of the type IIA superstring theory. 
In this limit, an approximation by the instanton number expansion does not make sense any longer. 
Instead there might be an appropriate description by weakly coupled degrees of freedom in an S-dual to the 
original theory. The behavior of the free energy might suggest the existence of such degrees of freedom. 

{}From the viewpoint of perturbation theory, the free energy is expected to be 
formally expressed 
as a double series with respect to $t^{-3/2}$ and $e^{-\frac{32}{3}\,t^{3/2}}$ 
(the so-called trans-series~\cite{Marino:2008ya}): 
\be
F_{(1,0)}=\sum_{k=1}^\infty e^{-\frac{32k}{3}\,t^{3/2}}\sum_{n=k}^\infty f^{(k)}_n\,t^{-\frac32n} 
\label{F10_doubleseries}
\ee 
with coefficients $f_n^{(k)}$. 
In matrix models for bosonic strings, 
it is extremely nontrivial to sum up such double series and obtain a well-defined result 
(for example, see~\cite{Marino:2008ya,David:1990sk,Chan:2010rw}). 
However, in our matrix model for the IIA superstring theory, 
Fig.~\ref{fig:FreeEnergy} indicates a well-defined result
after we manage the summation.

\section{Discussions}
\label{sec:discussions}
\setcounter{equation}{0}
In this paper, we explicitly computed nonperturbative effects in a supersymmetric double-well matrix model~\cite{Kuroki:2012nt} in the double scaling limit with $t= N^{2/3}\omega$ fixed.  
As was discussed in~\cite{Kuroki:2012nt,Kuroki:2013qpa}, 
this model corresponds to type IIA superstring theory on a nontrivial Ramond-Ramond 
background in the two-dimensional target space 
(Liouville direction) $\times$ ($S^1$ with self-dual radius). 

For the one-point function $\vev{\frac{1}{N}\tr\,(\phi^2-\mu^2)}^{(1,0)}$ 
and the free energy $F_{(1,0)}$, 
full one-instanton contributions were obtained as closed form expressions containing all 
perturbative fluctuations around the one-instanton background. 
Also, presented were their analytic expressions for the leading two-instanton effect 
with respect to finite but large $t$. 
The result shows that the supersymmetry is spontaneously broken by nonperturbative effects 
due to instantons. 
In particular, the instanton effects survive in the double scaling limit, which implies that 
supersymmetry breaking takes place by nonperturbative dynamics in the target space of 
the type IIA superstring theory. 

Moreover, we numerically evaluated full nonperturbative contributions to the one-point function and the free 
energy up to $N=1,000,000$, and extrapolated the results to $N=\infty$ in the double scaling limit. 
{}From the result, we further confirmed that $t$ is the correct scaling variable to be 
fixed in the double scaling limit. 
It was shown that the full one-instanton 
contribution to the one-point function gives a significantly better approximation of the $N=\infty$ result 
compared to the leading term in the one-instanton contribution, 
and that the leading two-instanton contribution further improves the result for $t\gtrsim 0.65$. 
The full nonperturbative free energy seems to be a finite function of $t$ even at the origin, which
corresponds to the strongly coupled limit of the type IIA superstring theory. 
The result might suggest  
a well-defined weakly coupled theory as an S-dual to the IIA theory. 
It would be intriguing to obtain an analytic expression for the full nonperturbative contribution 
and to identify the S-dual theory.

In order to identify the Nambu-Goldstone fermions associated with the spontaneous supersymmetry breaking, 
let us express the auxiliary field $B_{ij}$ by its expectation value 
\be
\vev{B_{ij}}^{(1,0)}=-i\vev{(\phi^2-\mu^2)_{ij}}^{(1,0)}
= -i\vev{\frac{1}{N}\,\tr(\phi^2-\mu^2)}^{(1,0)}\!\delta_{ij}
\ee
and fluctuations around it $\tilde{B}_{ij}$. The last equality comes from $U(N)$ symmetry of the system. 
Then, (\ref{QSUSY}) leads to a nonlinear transformation: 
\be
Q\bar{\psi}_{ij} = -\vev{\frac{1}{N}\,\tr(\phi^2-\mu^2)}^{(1,0)}\!\delta_{ij} -i\tilde{B}_{ij}, 
\ee
which is a signal of Nambu-Goldstone fermions according to the standard argument. 
Since the nonlinear term can be removed from $N-1$ of the $N$ independent components 
$\bar{\psi}_{ii}$ ($i=1,\cdots, N$) by taking appropriate linear combinations 
(for example, $\bar{\psi}_{11}-\bar{\psi}_{ii}$ ($i=2, \cdots, N$)), the linearly independent 
component $\frac{1}{N}\,\tr\,\bar{\psi}$ can be regarded as the Nambu-Goldstone fermion 
associated with the breaking of $Q$. 
Similarly, $\frac{1}{N}\,\tr\,\psi$ can be regarded as the Nambu-Goldstone fermion 
associated with the breaking of $\bar{Q}$. 
According to~\cite{Kuroki:2012nt,Kuroki:2013qpa}, 
these correspond to (R$+$, NS) and (NS, R$-$) vertex operators in the type IIA theory: 
\be
\int d^2z\, V_{+\frac12, \,+1}(z)\,\bar{T}_{\frac12}(\bar{z})=
\int d^2z\,e^{-\frac12\phi+\frac{i}{2}H+\frac{i}{2}x+\frac12\varphi}(z)\,
e^{-\bar{\phi}+\frac{i}{2}\bar{x} +\frac12\bar{\varphi}}(\bar{z})
\ee
and 
\be
\int d^2z\,T_{-\frac12}(z)\, \bar{V}_{-\frac12, \,-1}(\bar{z})=
\int d^2z\,e^{-\phi-\frac{i}{2}x+\frac12\varphi}(z)\,
e^{-\frac12\bar{\phi}-\frac{i}{2}\bar{H}-\frac{i}{2}\bar{x}+\frac12\bar{\varphi}}(\bar{z}) 
\ee
(up to cocycle factors), respectively. 
Note that the Nambu-Goldstone fermions are generally not identical to fermion zero-modes in an instanton background. 
Suppose that diagonalization of $\phi$: 
$\phi=U\,{\rm diag}(\lambda_1,\cdots, \lambda_N)\,U^\dagger$ ($U\in SU(N)$) is accompanied 
with transformations of the fermions: $\psi=U\psi'U^\dagger$ and $\bar{\psi}=U\bar{\psi}'U^\dagger$. 
As discussed in section~\ref{sec:MMinst}, one of the eigenvalues, say $y=\lambda_N$, sitting at the origin 
gives a configuration of a single instanton. Then, the fermionic variables $\psi'_{NN}$ and $\bar{\psi}'_{NN}$ 
will disappear from the classical action (\ref{S}), becoming fermion zero-modes. 
Likewise, for a $k$-instanton configuration in which $\lambda_j$ ($j=N-k+1, \cdots, N$) are at the origin, 
$2k^2$ fermionic variables $\psi'_{ij}$, $\bar{\psi}'_{ij}$ ($i,j=N-k+1,\cdots, N$) will become zero-modes. 
Almost all of these zero-modes would be lifted by quantum effects due to the Vandermonde determinant.  
It would be interesting to consider a relation between the Nambu-Goldstone fermions and the fermion 
zero-modes. 
The fermion zero-modes might suggest that each eigenvalue at the origin 
corresponds to an object like a D-brane and 
anti-D-brane pair in the type IIA theory. This would be clarified by considering analogs of 
FZZT or ZZ branes~\cite{Fateev:2000ik,Teschner:2000md,Zamolodchikov:2001ah} in the IIA superstring theory.

It would be interesting to consider various interpretations of the matrix variables in the matrix model. 
For instance, each matrix element could be regarded as a kind of string bit carrying 
a unit of winding or momentum along the $S^1$ target space, which seems to have some similarity to 
the matrix string theory~\cite{Dijkgraaf:1997vv}. 
Alternatively, the matrix elements might be interpreted as open string excitations 
on certain D-branes and the matrix model could describe closed string dynamics 
via the open-closed string duality~\cite{McGreevy:2003kb,Klebanov:2003km,McGreevy:2003ep,Takayanagi:2003sm,Douglas:2003up,McGreevy:2003dn}. 
Refs.~\cite{Seiberg:2004ei,Seiberg:2005bx,Mukherjee:2005aq,Kuroki:2006wn} would be useful 
for furthering an investigation along these lines.

\section*{Note added}
After this paper appeared in the arXiv, 
we were informed by Ricardo~Schiappa that refs.~\cite{Marino:2007te,Marino:2008vx,Schiappa:2013opa} 
generalize the method of \cite{Hanada:2004im,Ishibashi:2005dh} to be valid both in the double scaling limit 
and off-criticality. He also pointed out that trans-series and resurgent analysis discussed 
in~\cite{Pasquetti:2009jg,Aniceto:2011nu,Schiappa:2013opa} would be useful to compute higher instanton 
contributions in~(\ref{F10_doubleseries}). 
Mithat~\"{U}nsal informed us of trans-series and resurgence approach to nonperturbative completion of 
quantum field theory (for example, see~\cite{Dunne:2012ae,Dunne:2013ada,Cherman:2013yfa,Basar:2013eka}).

\section*{Acknowledgements}
We would like to thank Rajesh~Gopakumar, Hirotaka~Irie, Satoshi~Iso, Hiroshi~Itoyama, 
Shoichi~Kawamoto, Ivan~Kostov, Sanefumi~Moriyama, Koichi~Murakami, Ricardo~Schiappa, Hidehiko~Shimada, 
Yuji~Sugawara, Tsukasa~Tada, Tadashi Takayanagi, Mithat~\"{U}nsal  and Tamiaki~Yoneya 
for useful discussions and comments. 
The authors thank the Yukawa Institute for Theoretical Physics at Kyoto University and 
the High Energy Accelerator Research Organization (KEK). 
Discussions during the YITP workshop ``Gauge/Gravity duality'' (October, 2012) 
and the KEK theory workshop 2013 (March, 2013) were useful to complete this work. 
The work of M.~G.~E. is supported in part by MEXT Grant-in-Aid for Young Scientists (B), 23740227. 
The work of T.~K. is supported in part by Rikkyo University Special Fund for Research and 
JSPS Grant-in-Aid for Scientific Research (C), 25400274. 
The work of F.~S. is supported in part by JSPS Grant-in-Aid for Scientific Research (C), 
21540290 and 25400289. 
The work of H.~S. is supported in part by JSPS Grant-in-Aid for Scientific Research (C), 
23540330.

\appendix
\section{Perturbative contributions to $Z_{(1,0)}$}
\label{app:localization}
\setcounter{equation}{0}
In this appendix, we compute perturbative contributions to $Z_{(1,0)}$ by using a deformation method in 
topological field theory. Although related calculations are given in~\cite{Kuroki:2010au}, 
we present a direct derivation here. 
First, $Z_{(1,0)}$ in (\ref{Zff}) 
can be expressed in a form that involves the original matrix integrals:
\be
Z_{(1,0)}=\left(-1\right)^{N^2}
\int d^{N^2}B \int_{H_+}d^{N^2}\phi \int\left(d^{N^2}\psi \,d^{N^2}\bar{\psi}\right)\, e^{-S},  
\label{Z10_matrix}
\ee
with $S$ given by (\ref{S}). The integration region of $\phi$ is the space of positive definite 
Hermitian matrices $H_+$. 
We expand $\phi$ around the minimum $\mu$ of the double-well: 
\be
\phi= \mu + \tilde{\phi}, 
\ee
and compute $Z_{(1,0)}$ in perturbation theory with respect to small fluctuations $\tilde{\phi}$. 
The perturbative contribution can be expressed as 
\be
\left.Z_{(1,0)}\right|_{\rm pert.} \equiv \left(-1\right)^{N^2}
\int d^{N^2}B \,d^{N^2}\tilde{\phi} \,\left(d^{N^2}\psi \,d^{N^2}\bar{\psi}\right)\, e^{-S_{\rm free}}\,
\left[e^{-S_{\rm int.}}\right]_{\rm pert.}, 
\label{Z10_pert}
\ee
where 
\be
S_{\rm free} \equiv N\,\tr\left[\frac12B^2 +i2\mu B\tilde{\phi} +2\mu\bar{\psi}\psi\right], \qquad
S_{\rm int.} \equiv N\,\tr\left[iB\tilde{\phi}^2+\bar{\psi}\left(\tilde{\phi}\psi+\psi\tilde{\phi}\right)\right], 
\label{Sfree-Sint}
\ee
and the last factor $\left[e^{-S_{\rm int.}}\right]_{\rm pert.}$ means that 
the interaction part $S_{\rm int.}$ is treated in perturbation theory by expanding the exponential. 
Note that $\tilde{\phi}$ is integrated over all Hermitian matrices, following the conventional approach for 
perturbation theory around a saddle point of the classical action. 

Since the integrand of (\ref{Z10_pert}) is invariant under the supersymmetry transformations (\ref{QSUSY}) 
and (\ref{QbarSUSY}) with the trivial change of $\phi$ to $\tilde{\phi}$, 
adding the term $N\,\tr\, (\frac{\epsilon-1}{2}B^2)$ to the free part $S_{\rm free}$ does not affect 
the value of 
$\left.Z_{(1,0)}\right|_{\rm pert.}$ so long as the parameter $\epsilon$ is positive. 
For $\left.Z_{(1,0)}\right|_{\rm pert.}(\epsilon)$ denoting the deformed partition function, one can show 
\be
\frac{d}{d\epsilon}\left.Z_{(1,0)}\right|_{\rm pert.}(\epsilon)=0
\ee
for arbitrary positive $\epsilon$, from the facts that 
$N\,\tr \,(\frac{\epsilon-1}{2}B^2)$ can be written in a $Q$-exact or $\bar{Q}$-exact form
and that the deformation with $\epsilon>0$ does not change the asymptotic behavior of the integrand~\cite{Witten:1992xu}. 
Rescaling 
$B\to \frac{1}{\sqrt{\epsilon}}\,B$ and $\tilde{\phi}\to \sqrt{\epsilon}\,\tilde{\phi}$ 
after the deformation, we find the expression: 
\be
\left.Z_{(1,0)}\right|_{\rm pert.} = \left(-1\right)^{N^2}
\int d^{N^2}B \,d^{N^2}\tilde{\phi} \,\left(d^{N^2}\psi \,d^{N^2}\bar{\psi}\right)\, e^{-S_{\rm free}}\,
\left[e^{-\sqrt{\epsilon}\,S_{\rm int.}}\right]_{\rm pert.}. 
\label{Z10_pert2}
\ee
Because the value of $\left.Z_{(1,0)}\right|_{\rm pert.}$ does not depend on $\epsilon$, we may compute 
(\ref{Z10_pert2}) in the limit $\epsilon\to +0$.   
Then, the matrix integral reduces to trivial Gaussian integrations. 
By using (\ref{normalization1}) and (\ref{normalization2}), it is easy to obtain 
\be
\left.Z_{(1,0)}\right|_{\rm pert.} =1.
\label{Z10_pertf}
\ee
This result is valid for arbitrary $N$. 

\section{Computation of $V_{\rm eff}^{(1)}(0)$}
\label{app:Veff1}
\setcounter{equation}{0}
In this appendix, we compute the value at $y=0$ of 
the effective potential at the subleading order $V_{\rm eff}^{(1)}(y)$ in (\ref{Veff1}).  

\subsection{Computation of $\Delta_0D(y^2)$}
Let us first evaluate the contribution from the difference of disk amplitudes $\Delta_0D(y^2)$ 
defined by (\ref{DeltaDy2}). 
After the same replacement as (\ref{replacement_Z'}), 
$\vev{\frac{1}{N}\sum_{i=1}^{N-1}\frac{1}{z-\lambda_i^2}}_{\rm planar}^{'\,(1,0)}$ becomes 
\be
\sqrt{\frac{N-1}{N}}\,\cdot
\left.\vev{\frac{1}{N}\sum_{i=1}^{N}\frac{1}{z-\lambda_i^2}}_{\rm planar}^{(1,0)}\right|_{(N,\,\lambda_i,\,z,\,\mu)\to (N-1,\,\lambda'_i,\,z',\,\mu')},
\label{R2z_prime}
\ee
where $z=\left(\frac{N-1}{N}\right)^{\frac12}z'$. 
(\ref{R2z}) gives the last factor of (\ref{R2z_prime}) as~\footnote{
Note that the cut $[\mu'^2-2, \mu'^2+2]$ in the $z'$-plane implies the support of the eigenvalue distribution $[\sqrt{\mu'^2-2}, \sqrt{\mu'^2+2}]$. This is included in the integration region $[a', b']$.} 
\be
\frac12\left[z'-\mu'^2-\sqrt{(z'-\mu'^2)^2-4}\right].
\ee
Then, 
\bea
\vev{\frac{1}{N}\sum_{i=1}^{N-1}\frac{1}{z-\lambda_i^2}}_{\rm planar}^{'\,(1,0)} & = & 
\sqrt{\frac{N-1}{N}}\,\frac12\left[z'-\mu'^2-\sqrt{(z'-\mu'^2)^2-4}\right] \nn \\
 & = & \frac12\left[z-\mu^2-\sqrt{(z-\mu^2)^2-4\frac{N-1}{N}}\right] \nn \\
 & = & \vev{\frac{1}{N}\sum_{i=1}^{N}\frac{1}{z-\lambda_i^2}}^{(1,0)}_{\rm planar} 
-\frac{1}{N}\frac{1}{\sqrt{(z-a^2)(z-b^2)}} +\cO(N^{-2}). \nn \\
& & \label{deltaD_1}
\eea 
Similar to (\ref{trln_disk}), 
\be
\vev{\frac{1}{N}\sum_{i=1}^{N-1}\ln(y^2-\lambda_i^2)}_{\rm planar}^{'\,(1,0)} =\lim_{\Lambda\to\infty} 
\left[\int^{y^2}_\Lambda dz\vev{\frac{1}{N}\sum_{i=1}^{N-1}\frac{1}{z-\lambda_i^2}}^{'\,(1,0)}_{\rm planar}+\frac{N-1}{N}\ln \Lambda\right]. 
\label{trln_disk'}
\ee
Plugging (\ref{deltaD_1}) into (\ref{trln_disk'}), we have 
\be
\Delta_0D(y^2)=-2\lim_{\Lambda\to\infty}\left[{\rm Re}\int^{y^2}_\Lambda dz \frac{1}{\sqrt{(z-a^2)(z-b^2)}}
+\ln\Lambda\right],
\label{DeltaD_2}
\ee
whose value at the origin is computed as 
\bea
\Delta_0D(0) & = & 2\lim_{\Lambda\to\infty}\left[\int^\Lambda_{b^2}dz\frac{1}{\sqrt{(z-a^2)(z-b^2)}}
-\int^{a^2}_0dz\frac{1}{\sqrt{(a^2-z)(b^2-z)}}-\ln\Lambda\right] \nn \\
 & = & -\ln\frac{\mu^2+\sqrt{\mu^4-4}}{2}. 
\eea
The result turns out to be negligible in the double scaling limit: 
\be
\Delta_0D(0) = -2\frac{\sqrt{t}}{N^{1/3}}+\cO(N^{-2/3}).
\label{DeltaD_f}
\ee

\subsection{Computation of $\vev{\left\{{\rm Re}\sum_{i=1}^N\ln(y^2-\lambda_i^2)\right\}^2}_{C,\,{\rm planar}}^{(1,0)}$}
To obtain the annulus amplitude 
$\vev{\left\{{\rm Re}\sum_{i=1}^N\ln(y^2-\lambda_i^2)\right\}^2}_{C,\,{\rm planar}}^{(1,0)}$, 
let us start with the expression derived in appendix A of~\cite{Kuroki:2012nt}: 
\bea
& & \vev{\left(\sum_{i=1}^N\frac{1}{z-\lambda^2_i}\right)\left(\sum_{j=1}^N\frac{1}{w-\lambda^2_j}\right)}^{(1,0)}_{C,\,{\rm planar}} \nn \\
& & =\frac14\frac{1}{(z-w)^2}\left[\sqrt{\frac{(z-a^2)(w-b^2)}{(z-b^2)(w-a^2)}} 
+ \sqrt{\frac{(z-b^2)(w-a^2)}{(z-a^2)(w-b^2)}} -2\right],
\label{cyl_1}
\eea 
which is valid for an arbitrary filling fraction. 
Note that the r.h.s. can be written in terms of total derivatives as 
\be
\partial_z\partial_w\left[-\ln\left(1-e^{-\theta_z}e^{-\theta_w}\right)\right]
\label{cyl_2}
\ee
with 
\be
e^{-\theta_z}=\vev{\frac{1}{N}\sum_{i=1}^N\frac{1}{z-\lambda_i^2}}^{(1,0)}_{\rm planar}=
\frac12\left[z-\mu^2-\sqrt{(z-a^2)(z-b^2)}\right].
\ee
Also, 
\be
e^{\pm\frac12\theta_z}=\frac12\left(\sqrt{z-a^2}\pm\sqrt{z-b^2}\right)
\ee 
will be useful to check (\ref{cyl_2}). 

Similar to (\ref{trln_disk}), we have 
\bea
& & \vev{\left(\sum_{i=1}^N\ln(z-\lambda_i^2)\right)\left(\sum_{j=1}^N\ln(w-\lambda_j^2)\right)}^{(1,0)}_{C,\,{\rm planar}} \nn \\
& & = \int^z_\infty dz'\int^w_\infty dw'\vev{\left(\sum_{i=1}^N\frac{1}{z'-\lambda^2_i}\right)\left(\sum_{j=1}^N\frac{1}{w'-\lambda^2_j}\right)}^{(1,0)}_{C,\,{\rm planar}}, 
\label{trln_cyl}
\eea
where we have used the fact that the leading terms ($\ln z$ and $\frac{1}{z}$) in the large-$z$ expansions of 
$\ln(z-\lambda_i^2)$ and $\frac{1}{z-\lambda_i^2}$ do not contribute to the connected amplitudes. 
Plugging (\ref{cyl_2}) into (\ref{trln_cyl}) leads to 
\be
\vev{\left(\sum_{i=1}^N\ln(z-\lambda_i^2)\right)\left(\sum_{j=1}^N\ln(w-\lambda_j^2)\right)}^{(1,0)}_{C,\,{\rm planar}} =- \ln\left(1-e^{-\theta_z}e^{-\theta_w}\right).
\ee
Thus, for $|y|>b$, 
\bea
\vev{\left\{{\rm Re}\sum_{i=1}^N\ln(y^2-\lambda_i^2)\right\}^2}_{C,\,{\rm planar}}^{(1,0)}
& = & \vev{\left\{\sum_{i=1}^N\ln(y^2-\lambda_i^2)\right\}^2}_{C,\,{\rm planar}}^{(1,0)} \nn \\
& = & -\ln\left[1-\frac14\left\{y^2-\mu^2-\sqrt{(y^2-a^2)(y^2-b^2)}\right\}^2\right]. \nn \\
& & 
\eea
For $|y|<a$, although
\be
\ln(y^2-\lambda_i^2)= {\rm Re}\ln(y^2-\lambda_i^2)\pm i\pi,
\ee
the argument $\pm i\pi$ is constant and does not contribute to the connected amplitude. 
We obtain 
\bea
\vev{\left\{{\rm Re}\sum_{i=1}^N\ln(y^2-\lambda_i^2)\right\}^2}_{C,\,{\rm planar}}^{(1,0)}
= -\ln\left[1-\frac14\left\{y^2-\mu^2+\sqrt{(a^2-y^2)(b^2-y^2)}\right\}^2\right],  \nn \\
& & 
\eea
whose value at the origin becomes 
\bea
\left.\vev{\left\{{\rm Re}\sum_{i=1}^N\ln(y^2-\lambda_i^2)\right\}^2}_{C,\,{\rm planar}}^{(1,0)}\right|_{y=0} 
& = & -\ln\left[1-\left(\frac{b-a}{2}\right)^4\right] \nn \\
& = & -\ln\frac{4\sqrt{t}}{N^{1/3}}+\cO(N^{-1/3}) 
\label{trln_cyl_f}
\eea
in the double scaling limit. 

\subsection{Result of $V_{\rm eff}^{(1)}(0)$}
{}From (\ref{DeltaD_f}) and (\ref{trln_cyl_f}), the subleading order contribution 
to the effective potential at the origin 
is obtained as 
\be
V_{\rm eff}^{(1)}(0) = \ln\frac{16t}{N^{2/3}}+\cO(N^{-1/3}). 
\label{app:Veff1_f}
\ee

\section{Derivation of (\ref{hermite_asymp})}
\label{app:formula}
\setcounter{equation}{0}
The formula (\ref{hermite_asymp}) with an unspecified  $\cO(n^{-2/3})$ correction is found in eq.~(8.22.14) of ref.~\cite{szego}. 
In order to make this paper self-contained, 
we derive the formula (\ref{hermite_asymp}) and explicitly determine the $\cO(n^{-2/3})$ correction $g_0(s)$ 
which is required to obtain (\ref{Omega1_2/3}). 

Although (\ref{hermite_asymp}) seems to be understood as contributions around the turning points in the WKB analysis of 
the harmonic oscillator potential, let us start with the integral representation 
\be
H_n(x) = \frac{2^n}{\sqrt{\pi}}\int^\infty_{-\infty}dt\,e^{-t^2}\,(x+it)^n  
\ee
for a more systematic treatment. 
At the value of $x$ in (\ref{x_s}), it takes the form 
\be
H_n(x) = \frac{2^n}{\sqrt{\pi}}\,(2n+1)^{\frac{n+1}{2}}\int^\infty_{-\infty}dt\, e^{f(t)} 
\label{hermite_int}
\ee
with
\be
f(t) \equiv -(2n+1)t^2+n\,\ln (1+\Delta +it), \qquad \Delta \equiv \frac{s}{\sqrt{2}\,n^{1/6}\sqrt{2n+1}}.
\ee 

We evaluate the integral (\ref{hermite_int}) using the saddle point method for $n$ large. 
The saddle point satisfying $f'(t)=0$ and $f''(t)< 0$ is 
\be
t_-=\frac{i}{2}\left[1+\Delta -\sqrt{2\Delta+\Delta^2+\frac{1}{2n+1}}\right].
\ee
$f(t)$ must be expanded around the saddle point up to fifth order to obtain 
$g_0(s)$ in (\ref{hermite_asymp}): 
\be
f(t)=f(t_-) + \frac{1}{2!}f''(t_-)\frac{z^2}{n^{2/3}}+ \frac{1}{3!}f'''(t_-)\frac{z^3}{n} 
+\frac{1}{4!}f^{(4)}(t_-)\frac{z^4}{n^{4/3}}+ \frac{1}{5!}f^{(5)}(t_-)\frac{z^5}{n^{5/3}} +\cdots.
\ee
Here, we take $t-t_-=\frac{z}{n^{1/3}}$. This choice naturally follows from the fact that 
$f''(t_-)$ is a quantity of order $n^{2/3}$. 
Explicitly, 
\bea
f(t) & = & f(t_-) -4s^{1/2}z^2-i\frac83 z^3 +n^{-1/3}\left\{(4s-s^{-1/2})z^2+i8s^{1/2}z^3-4z^4\right\} \nn \\
 &  & +n^{-2/3}\left\{\left(-\frac52s^{3/2}+2+\frac18 s^{-3/2}\right)z^2+i2(s^{-1/2}-6s)z^3 +16s^{1/2}z^4+i\frac{32}{5}z^5
 \right\} \nn \\
 & & +\cO(n^{-1}), \\
f(t_-) & = & \frac{n}{2}-n\ln 2 + n^{1/3}s -\frac23 s^{3/2} +\frac14 
+ n^{-1/3}\left(-\frac12 s^{1/2}+\frac14s^2\right) \nn \\
& & + n^{-2/3}\left(-\frac{1}{16}s^{-1/2}+\frac14 s-\frac{1}{20}s^{5/2}\right)+ \cO(n^{-1}).
\eea

Under the variable change $z \to \frac12(z+is^{1/2})$, the integral (\ref{hermite_int}) becomes 
\be
H_n(x) = \frac{2^n}{\sqrt{\pi}}\frac{(2n+1)^{\frac{n+1}{2}}}{2n^{1/3}}\,e^{f(t_-)+\frac23 s^{3/2}} 
\int^\infty_{-\infty}dz \, e^{-isz-i\frac13 z^3}\,e^{A(s,z)}.
\ee 
The integral gives the Airy function up to the small correction terms attributed to   
\bea
A(s,z) & \equiv & n^{-1/3}\left\{\frac14(s^{1/2}-s^2)-\frac{i}{2}z-\left(\frac12s+\frac14s^{-1/2}\right)z^2-\frac14z^4\right\} \nn\\
 & & +n^{-2/3}\left\{-\frac{1}{32}s^{-1/2}-\frac14s-\frac{3}{40}s^{5/2}+i\frac{(1+2s^{3/2})^2}{16s}z 
\right. \nn \\
 & & \left. \hspace{14mm} +\frac{1}{32}(-8+s^{-3/2}-4s^{3/2})z^2 
+i\left(\frac14s^{-1/2}+\frac12s\right)z^3+i\frac15z^5\right\} \nn \\
& & +\cO(n^{-1}).
\eea  

The l.h.s. of (\ref{hermite_asymp}) is written as 
\be
e^{-x^2/2}H_n(x) = \pi^{\frac14} 2^{\frac{n}{2}+\frac14} n^{-\frac{1}{12}}\sqrt{n!}\,e^{B(s)}
 \frac{1}{2\pi}\int^\infty_{-\infty}dz\,e^{-isz-i\frac13 z^3}\,e^{A(s,z)}
\label{hermite_asymp1} 
\ee
with 
\be
B(s)\equiv -\frac12 n^{-1/3}s^{1/2} -n^{-2/3}\left(\frac{1}{16}s^{-1/2}+\frac{1}{20}s^{5/2}\right) 
+\cO(n^{-1}).
\ee 
In (\ref{hermite_asymp1}), we keep terms up to $\cO(n^{-2/3})$ in the expansion 
$e^{B(s)+A(s,z)}=1+B(s)+A(s,z)+\frac12 (B(s)+A(s,z))^2+ \cdots$. These terms can be removed from the integral 
by replacing $z$ with $i\partial_s$. 
Then, 
\bea
e^{-x^2/2}H_n(x) & = & \pi^{\frac14} 2^{\frac{n}{2}+\frac14} n^{-\frac{1}{12}}\sqrt{n!} \nn \\ 
& & \times\left[1+n^{-1/3}C_1(s,\partial_s)+ n^{-2/3}C_2(s,\partial_s)+\cO(n^{-1})\right]{\rm Ai}(s),  
\eea
where
\bea
C_1(s,\partial_s) & \equiv & -\frac14(s^{1/2}+s^2)+\frac12\partial_s+\left(\frac12s+\frac14s^{-1/2}\right)\partial_s^2-\frac14\partial_s^4, \\
C_2(s,\partial_s) & \equiv & -\frac{3}{32}s^{-1/2}-\frac{7}{32}s-\frac{s^{5/2}}{16}+\frac{s^4}{32}
-\left(\frac{1}{16}s^{-1}+\frac38(s^{1/2}+s^2)\right)\partial_s \nn \\
& & +\left(\frac{5}{16}-\frac{1}{32}s^{-3/2}-\frac{s^{3/2}}{16}-\frac{s^3}{8}\right)\partial_s^2 
+\frac34\left(\frac12s^{-1/2}+s\right)\partial_s^3 \nn \\
& & +\left(\frac{1}{32}s^{-1}+\frac{3}{16}(s^{1/2}+s^2)\right)\partial_s^4-\frac{13}{40}\partial_s^5 
-\frac18\left(\frac12s^{-1/2}+s\right)\partial_s^6 \nn \\
& & +\frac{1}{32}\partial_s^8 .
\eea
Use of 
\bea
\partial_s^2{\rm Ai}(s)& = & s{\rm Ai}(s), \nn \\
\partial_s^3{\rm Ai}(s)& = & (1+s\partial_s){\rm Ai}(s), \nn \\
\partial_s^4{\rm Ai}(s)& = & (s^2+2\partial_s){\rm Ai}(s), \nn \\
\partial_s^5{\rm Ai}(s)& = & (4s+s^2\partial_s){\rm Ai}(s), \nn \\
\partial_s^6{\rm Ai}(s)& = & (s^3+4+6s\partial_s){\rm Ai}(s), \nn \\
\partial_s^8{\rm Ai}(s)& = & (s^4+28s+12s^2\partial_s){\rm Ai}(s) 
\eea
leads to (\ref{hermite_asymp}) with 
\be
g_0(s)=-\frac{s}{20}\,{\rm Ai}(s) + \frac{s^2}{20}\,{\rm Ai}'(s).
\label{g0}
\ee

\section{Evaluation of $\Omega(N, t)$ at the subleading of $1/N$}
\label{app:1ptfn_subleadingN}
\setcounter{equation}{0}
%
\begin{figure}
\centering
\includegraphics[height=7cm, width=12cm, clip]{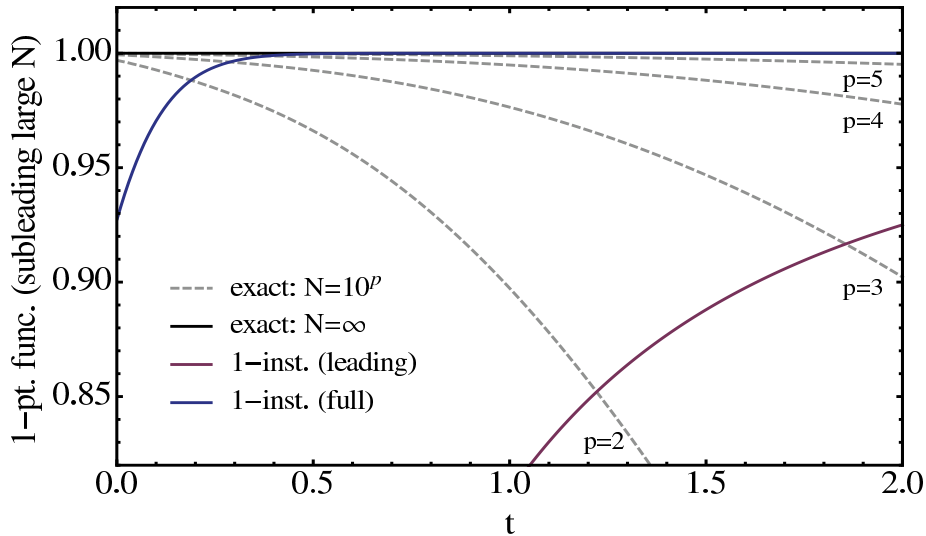}
\caption{The subleading in $1/N$ contribution 
to the one-point function $\Omega(N, t)$ 
as a function of $t$. 
Everything is normalized by the extrapolation of 
(\ref{1ptfn_subleading_finiteN}) to $N=\infty$ (exact: $N=\infty$), and thus 
the black solid line representing it is flat. The gray dashed lines (exact: $N=10^p$) 
show (\ref{1ptfn_subleading_finiteN}) for $N=10^p$ ($p=2, 3, 4, 5$). 
(\ref{Omega1_2/3}) is depicted by the blue line (1-inst. (full)), 
and the leading order term of (\ref{Omega1_2/3}) at large $t$ by the red line (1-inst. (leading)). } 
\label{fig:1ptfn_subleadingN}
\end{figure}
%
In Fig.~\ref{fig:1ptfn_subleadingN}, 
we present a plot of the subleading large-$N$ contribution 
to the one-point function 
$\Omega(N, t)$ 
in the double scaling limit. 
The result is normalized by the full nonperturbative result obtained from numerical extrapolation of 
the difference 
\be
N^{2/3}\left(\Omega(N,t)-\Omega_0(t)\right),  
\label{1ptfn_subleading_finiteN}
\ee
evaluated at various finite $N$ (up to $N=1,000,000$). 
For comparison, we also present the result for the full one-instanton contribution (\ref{Omega1_2/3}) and its 
leading order contribution at large $t$. 
In the plot, it appears that even the subleading large-$N$ contribution to 
the full one-instanton result (\ref{Omega1_2/3}) 
exhibits good agreement with the numerical results 
for the full contribution 
for $t\gtrsim 0.5$.


\end{document}